\documentclass[aps,a4paper, preprint, superscriptaddress,preprintnumbers,floatfix,nofootinbib,amsmath,amssymb]{revtex4}
\usepackage{url}
\usepackage{hyperref}
\usepackage{cleveref}

\usepackage{amstext,amssymb}
\usepackage{amsmath}
\usepackage{graphicx}
\usepackage{xspace}
\usepackage{color}
\usepackage{units}
\usepackage{slashed} 
\usepackage{multirow}
\newcommand{\be}{\begin{equation}}
\newcommand{\ee}{\end{equation}}
\newcommand{\bea}{\begin{eqnarray}}
\newcommand{\eea}{\end{eqnarray}}
\newcommand{\nn}{\nonumber}

\def\s1{\hat s}

\usepackage{slashed}

\newcommand{\nua}[1]{\ensuremath{\rlap{\kern-2.5pt\ensuremath{\overset{\scriptscriptstyle(-)}{\phantom{\nu}}}}{\ensuremath{{\nu}_{#1}}}}\xspace}

\newcommand{\eVq}  {\text{eV}^2}

\begin{document}
\title{ Neutrino Phenomenology and Dark matter in an $A_4$ flavour extended $B-L$ model}
\author{Subhasmita Mishra}
\email{subhasmita.mishra92@gmail.com}
\affiliation{Department of Physics, IIT Hyderabad, Kandi - 502285, India}
\author{ Mitesh Kumar Behera}
\email{miteshbehera1304@gmail.com}
\affiliation{School of Physics,  University of Hyderabad, Hyderabad - 500046,  India}
\author{ Rukmani Mohanta}
\email{rmsp@uohyd.ac.in}
\affiliation{School of Physics,  University of Hyderabad, Hyderabad - 500046,  India}
\author{ Sudhanwa Patra}
\email{sudhanwa@iitbhilai.ac.in}
\affiliation{Indian Institute of Technology Bhilai, GEC Campus, Sejbahar, Raipur-492015, India}
\author{ Shivaramakrishna Singirala}
\email{krishnas542@gmail.com}
\affiliation{Discipline of Physics, Indian Institute of Technology Indore, Indore-453 552, India}

\begin{abstract}
We present an $\rm A_4$ flavor extended $\rm B-L$ model for realization of eV scale sterile neutrinos, motivated by the recent experimental hints from both particle 
physics and cosmology. The framework considered here  is a gauged $\rm B-L$ extension of standard model without the introduction of right-handed neutrinos, where the gauge triangle  anomalies are canceled with the inclusion of  three exotic neutral fermions $N_{i}$  ($i=1,2,3$)  with $\rm B-L$ charges $-4,-4$ and $5$. The usual Dirac Yukawa couplings between the SM neutrinos and the 
exotic fermions  are absent and thus, the model allows natural realization of eV scale sterile-like neutrino and its mixing with standard model neutrinos by invoking $\rm A_4$ flavor symmetry. We demonstrate how the exact tri-bimaximal mixing pattern is perturbed due to active-sterile mixing by analyzing 
$1+3$ case in detail.  
We also show the implication of  eV scale sterile-like neutrino  on various observables in  
neutrino oscillation experiments and the effective mass in neutrinoless double beta decay.  Another interesting feature of the model is that one of three exotic fermions is required to explain eV scale phenomena, while other two fermions form stable dark matter candidates and their total relic density satisfy the observed $3\sigma$ limit of Planck data.
We constrain the gauge parameters associated with $U(1)$ gauge extension, using relic density and collider bounds.
\end{abstract}

\maketitle
\flushbottom

\section{Introduction} 
Albeit its success, Standard Model (SM) is not the  complete theory of nature to explain many observed phenomena. 
Neutrino oscillation experiments, in contrast to the zero mass prediction of SM, have confirmed the need for massive neutrinos and thus, necessitates for physics beyond the SM (BSM).  
The massive neutrinos and most of neutrino oscillation data can be explained in a framework of three active neutrinos through the elegant canonical seesaw mechanism \cite{Mohapatra:1979ia,Barr:2003nn, Akhmedov:2006de,Akhmedov:1999qe,Cai:2017jrq,Miranda:2016ptb},
whereas some experimental observations are strongly hinting towards one or two additional light neutrinos with ${\rm eV}$ scale masses and $\mathcal{O}(0.1)$ mixing with 
active neutrinos~\cite{Peres:2000ic,Sorel:2003hf,Vagnozzi:2017ovm,Kopp:2011qd,Giunti:2011gz,Bonilla:2018ynb,Heeck:2012zb}, stemming from particle physics, cosmology and astrophysics (for details, reader may 
refer to the white paper~\cite{Aguilar:2001ty}). While large number of experiments are coming up in next few years in order to investigate the possible presence of eV scale sterile neutrinos, which would be a ground-breaking discovery, there are few model building efforts in this direction. The aim of this work is to provide a simple BSM framework explaining the presence of  one eV scale sterile neutrino along with  ${\cal O}(0.1)$ mixing with active neutrinos and their effects on
neutrinoless double beta decay (NDBD). This model also provides a detail study of DM phenomenology via a TeV scale fermionic DM and the collider constraints from $Z^\prime$ mass. 

The smallness of neutrino mass and their hierarchical structure become one of the most challenging problems in particle physics. In the standard scenario of three active neutrino oscillation, two mass-squared differences of order $10^{-5} ~\rm eV^2$ and  $10^{-3} ~\rm eV^2$,  are observed from solar and atmospheric neutrino oscillation experiments respectively  \cite{Tanabashi:2018oca}. In fact, the absolute scale of neutrino mass still remains as an open question to be solved, however, there exists an upper bound on the sum  of active neutrino masses, $\sum m_\nu < 0.12 $ eV from  cosmological observations \cite{Aghanim:2018eyx}. Over the last two decades, 
several dedicated experiments have determined the neutrino  oscillation parameters rather precisely, though  few of them are still unknown. These include the neutrino mass ordering, the exact domain of the atmospheric mixing angle $\theta_{23}$ (octant problem) and the CP violating phase $\delta_{\rm CP}$.
%
But there exist few experimental anomalies, which can not be explained within the standard three neutrino framework. Of them, one is  the possible presence of sterile neutrinos  \cite{Blanco:2019vyp}, which is evident from the anti-neutrino flux measurement in LSND \cite{Aguilar:2001ty} and MiniBooNE \cite{Aguilar-Arevalo:2013pmq} experiments. The excess flux of $\bar{\nu}_e$ in appearance mode during $\nu_\mu \rightarrow \nu_e$ oscillation, hints towards the possible existence of at least one additional state  with  eV scale mass \cite{Dentler:2018sju}. This new state should not have any gauge interaction as per the $Z$-boson precision measurement and hence being sterile in nature. 
Thus, theoretical explanation of eV scale sterile neutrinos and its order of $0.1$ mixing with the active neutrinos is worth to study through possible BSM frameworks. 
The non-trivial mixing of such a light sterile neutrino with SM neutrinos are well studied in the literature within different seesaw framework~\cite{Barry:2011wb,
Merle:2011yv,Zhang:2011vh,Dev:2012bd,Zhang:2013ama,Das:2017ski,Das:2019fee,Frank:2013wga,Adhikari:2015woo,Nath:2016mts,Das:2018qyt}. Though there seems to be  a tension between the cosmological observation and neutrino oscillation results from  short baseline experiments regarding the existence of eV scale sterile neutrino, which is quite incompatible with the cosmologically observed value of $N_{eff}$, there exist literature, where different solutions to these issues are discussed \cite{Gariazzo:2019gyi,Pires:2019elj,Hannestad:2013ana,Archidiacono:2015oma,Dasgupta:2013zpn,Hannestad:2012ky}. However, in our model we are not discussing a detailed study of $N_{eff}$, well keeping it as an interesting scope for the follow up work of the current model. Along with these issues, the nature of neutrinos, 
i.e., whether it is  Dirac or Majorana also remains unexplained. The only way to test the Majorana nature of neutrinos is through the extremely rare lepton-number violating neutrino-less double beta decay (NDBD)
 experiments.

One more well-known challenging problem in particle cosmology  is that, SM doesn't have any explanation about the existence  of DM even though 
we have enough indirect gravitational evidence about its existence. Attempts have been made through DM getting scattered off the SM particles i.e., in the context of direct searches, and the well known collaborations include LUX, XENON, PICO, PandaX etc \cite{Akerib:2013tjd,Aprile:2012nq}. Study of excess in positron, electron or Gamma excess i.e., indirect signals, and the experiments include AMS-02, H.E.S.S, MAGIC, Fermi-LAT etc \cite{Aartsen:2012kia,Ackermann:2013uma,Aguilar:2013qda,Abramowski:2013ax,Aleksic:2013xea}. Apart from these, dark sector particle production is also probed in the accelerator experiments as well.
    
   To explain these experimental discrepancies, SM needs to be extended with extra symmetries or particles. Discrete symmetries are mostly preferred by the phenomenologists for  model building purpose as they restrict  the interaction terms by giving a specific structure to the mass matrix. $\rm A_4$ flavor symmetry is widely used in neutrino phenomenology as it gives the simple tribimaximal (TBM) mixing, which is more or less compatible with the standard neutrino mixing matrix ($\rm U_{PMNS}$). But this mixing pattern predicts a vanishing reactor mixing angle $\theta_{13}$ \cite{Petcov:2018mvm}, which conflicts the current experimental observation. To address this issue various ways are preferred to perturb the TBM structure by adding minimal number of flavon fields and myriad amount of literature is focused on neutrino phenomenology with $\rm A_4$ symmetry in the frameworks of different seesaw mechanism \cite{Ma:2001dn,Altarelli:2005yx,Altarelli:2009kr,Petcov:2018snn,Ma:2012ez,King:2013eh,Petcov:2018mvm}. Apart from neutrino phenomenology, phenomenological study of DM has been made within $\rm A_4$ framework \cite{Boucenna:2011tj,Meloni:2010sk,Mukherjee:2015axj,Bhattacharya:2016lts,delaVega:2018cnx}, but very few literature have been devoted to study these phenomena with gauge extended $\rm A_4$ flavor symmetric model. We consider a minimal extension of SM with $\rm A_4$ and $\rm U(1)_{B-L}$ symmetry in addition to three flavon fields and three singlet scalars, responsible for the breaking of $\rm A_4$ and $\rm U(1)_{B-L}$ symmetry respectively. This model includes three additional fermions with exotic $\rm B-L$ charges of $-4$, $-4$,  $5$ to protect from triangle gauge anomalies. This extension enhances the predictability of the model by explaining different phenomenological consequences like DM, neutrino mass and NDBD, compatible with the current observations. The minimal $B-L$ gauge extension of Standard Model with one copy of right-handed neutino having $B-L$ charge as $-1$ per generation or popular left-right theories~\cite{Mohapatra:1974gc, Pati:1974yy, Senjanovic:1975rk, Senjanovic:1978ev,
Mohapatra:1980yp} (where $B-L$ charge has a physical meaning with electric charge) provides an easier platform to explain the light neutrino masses through seesaw mechanism as well as the baryon asymmetry of the universe through leptogenesis. However, these theories fail to accomodate a stable cold dark matter candidate without imposition of ad-hoc discrete symmetry or inclusion of additional particles (see a recent work~\cite{Heeck:2015qra} for minimal left-right dark matter). The origin of neutrino mass and leptogenesis in these frameworks have been governed by a Dirac type Yukawa interaction term $y_\nu \overline{\ell_L} H N_R$. But the present $B-L$ extension of Standard Model with extra neutral fermions having exotic choice of $B-L$ charges can have  potential stable cold dark matter candidates.  The stability of dark matter is ensured by forbidding the generic interactions with SM leptons at renormalizable level and thereby, it can provide interesting dark matter phenomenology in both scalar and new gauge portal sector.
 
           This manuscript is structured as: section 2 follows the brief description of model and particle content along with the full Lagrangian and symmetry breaking. In section 3, we discuss the neutrino masses and mixing with one sterile-like neutrino scenario and section 4 includes the contribution of active-sterile mixing to the NDBD as per current experimental observation. In section 5, we illustrate a detailed description of DM phenomenology and collider bounds on new gauge parameters within the model framework. In section 6, we summarize and conclude the phenomenological consequences of the model.

\section{Model Description}
We propose a new variant of $\rm U(1)_{B-L}$ gauge extension of SM with additional $\rm A_4$ flavor symmetry, which includes three new neutral fermions $N_i$'s ($i = 1,2,3$) with exotic $\rm B-L$ charges $-4,-4$ and $+5$ to nullify the triangle gauge anomalies \cite{Montero:2007cd}. This choice of adding three exotic fermions in the context of $\rm B-L$ framework has been explored in several previous works \cite{Ma:2014qra,Ma:2015raa,Singirala:2017see,Singirala:2017cch,Geng:2017foe,Nomura:2017jxb,Nomura:2017vzp,Nomura:2017kih}. The spontaneous symmetry breaking of $\rm B-L$ gauge symmetry is realized by assigning non-zero vacuum expectation value (VEV) to three singlet scalars ($\phi_{2}, \phi_4$ and $\phi_8$), which also generate mass terms to all exotic fermions and the new gauge boson. Additionally, $\rm A_4$ flavor symmetry is used to study the neutrino phenomenology in this model. Apart from the usual SM Higgs doublet and the above mentioned three scalar singlets, which are responsible for electroweak and $\rm U(1)_{B-L}$ gauge symmetry breaking respectively, there are three SM singlet flavon fields $\phi_T$, $\chi$, $\zeta$ to break the $\rm A_4$ flavor symmetry at high scale. In the current framework, we chose the VEV of flavon fields to be much higher than the VEV of $\phi_2$, $\phi_4$ and $\phi_8$ to explain the neutrino phenomenology. Finally we impose a $Z_3$ symmetry to forbid the unwanted terms in the interaction Lagrangian. 

In this work, we intend to provide a detailed description of oscillation phenomenology with one sterile-like neutrino scenario. The complete field content with their corresponding charges are provided in Tables \ref{model_SM} and \ref{model_new}. The multiplication rules under $\rm A_4$ symmetry group is outlined in Appendix. 
\begin{table}[htb]
\centering
\begin{tabular}{|c|c|c|c|c|c|c|c|c|c|}
\hline
Field ~&~$L$ ~&~$e_R$~&~ $\mu_R$ ~&~ $\tau_R$  ~&~ $H$      \\
\hline
\hline
$\rm SU(2)_L$~&~$2$~&~$1$~ &~$1$~&~$1$~&~  $2$       \\
\hline
$\rm A_4$~&~$3$~&~$1$&$1''$~&~$1'$~&~$1$                \\
\hline
$\rm U(1)_{B-L}$~&~$-1$~&~$-1$&$-1$~&~$-1$~&~$0$     \\
\hline
$Z_3$~&~$1$~&~$\omega^2$&$\omega^2$~&~$\omega^2$~&~$1$     \\
\hline
\end{tabular}
\caption{SM field content of lepton and Higgs sectors alongwith their corresponding charges.}\label{model_SM}
\end{table}
\begin{table}[htb]
\centering
\begin{tabular}{|c|c|c|c|c|c|c|c|c|c|}
\hline
Field               & $N_{1}$~&~$N_{2}$  ~&~ $N_{3}$ &  $\phi_2$     & $\phi_4$    & $\phi_{8}$      & $\phi_T$    & $\chi$   & $\zeta$    \\
\hline
\hline
$\rm SU(2)_L$    & $1$ & $1$ & $1$ & $1$              & $1$            & 1                    & $1$            & $1$      &  $1$         \\
\hline
$\rm A_4$          & $1$ & $1$ & $1$   & $1$              &  $1$           & 1                    &  $3$           & $3$      &  $1'$         \\
\hline
$\rm U(1)_{B-L}$  & $5$ & $-4$ & $-4$ &$2$             &  $4$           & 8                    &   0        &0      &    0         \\
\hline
$Z_3$  & $1$ & $1$ & $1$ & $1$             &  $1$           & $1$                    & $\omega$        &$1$      &    $1$       \\
\hline
\end{tabular}
\caption{Complete field content with their corresponding charges of the proposed model.}\label{model_new}
\end{table}
\subsection{Scalar potential and symmetry breaking pattern}
Considering all the scalar content of the model, the  potential can be written as
 \begin{eqnarray}
V &=& \mu^2_{ H}  (H^\dagger H) + \lambda_{ H} (H^\dagger H)^2 + \mu^2_2 (\phi^\dagger_2 \phi_2) + \lambda_{22} (\phi^\dagger_2 \phi_2)^2 + \mu^2_4 ( \phi^\dagger_4 \phi_4)  + \lambda_4 (\phi^\dagger_4 \phi_4)^2\nonumber \\
     &&+\mu^2_8(\phi_8^\dagger \phi_8)+\lambda_8(\phi_8^\dagger \phi_8)^2 + \mu^2_{T} (\phi_{T}^\dagger {\phi_{T}}) + \lambda_{T} (\phi_{T}^\dagger {\phi_T})^2+\mu_\chi^2(\chi^\dagger \chi)+\lambda_\chi(\chi^\dagger \chi)^2  \nn\\
     &&+\mu_\zeta^2(\zeta^\dagger \zeta)+\lambda_\zeta(\zeta^\dagger \zeta)^2+\lambda_{\rm H2} (H^\dagger H) (\phi^\dagger_2 \phi_2)+\lambda_{\rm H4} (H^\dagger H) (\phi^\dagger_4 \phi_4)+\lambda_{\rm H\chi} (H^\dagger H) (\chi^\dagger \chi)\nn\\
    &&+\lambda_{ H\zeta} (H^\dagger H) (\zeta^\dagger \zeta)+\lambda_{H8}(H^\dagger H)(\phi_8^\dagger \phi_8)+\lambda_{\rm HT} (H^\dagger H) (\phi^\dagger_T \phi_T) + \lambda_{24} (\phi^\dagger_2 \phi_2) (\phi^\dagger_4 \phi_4) \nonumber \nn\\
      &&+\lambda_{28}(\phi_2^\dagger \phi_2)(\phi_8^\dagger \phi_8)+\lambda_{2T} (\phi_{T}^\dagger \phi_{T}) (\phi^\dagger_2 \phi_2)+\lambda_{ 2\zeta} (\phi^\dagger_2 \phi_2) (\zeta^\dagger \zeta)+\lambda_{2\chi} (\phi^\dagger_2 \phi_2) (\chi^\dagger \chi) \nn\\
 &&+\lambda_{48}(\phi_4^\dagger \phi_4)(\phi_8^\dagger \phi_8)+ \lambda_{ 4T} (\phi_{ T}^\dagger \phi_{\rm T}) (\phi^\dagger_4 \phi_4)+ \lambda_{\rm 4 \chi} (\phi_{ 4}^\dagger \phi_{4}) (\chi^\dagger \chi)+\lambda_{ 4 \zeta} (\phi_{ 4}^\dagger \phi_{4}) (\zeta^\dagger \zeta) \nn \\
&&+\lambda_{8T}(\phi_8^\dagger \phi_8)(\phi_T^\dagger \phi_T)+\lambda_{8\chi}(\phi_8^\dagger \phi_8)({\chi}^\dagger \chi)+\lambda_{8\zeta}(\phi_8^\dagger \phi_8)({\zeta}^\dagger \zeta)+ \lambda_{\chi T}(\chi^\dagger \chi)(\phi_{ T}^\dagger \phi_{T}) \nn \\
&&+\lambda_{\chi \zeta }(\chi^\dagger \chi)({\zeta}^\dagger \zeta)+\lambda_{\rm \zeta T}(\zeta^\dagger \zeta)(\phi_T^\dagger \phi_T) +\mu_{24} \left((\phi_2)^2 \phi_4^\dagger + (\phi_2^\dagger)^2 \phi_4\right)\nn \\
&& +\lambda_{248}\left((\phi_2)^2\phi_4 \phi_8^\dagger +(\phi_2^\dagger)^2\phi_4^\dagger {\phi_8} \right)+\mu_{48}\left((\phi_4)^2\phi_8^\dagger +(\phi_4^\dagger )^2{\phi_8}\right).
\label{scalar}
\end{eqnarray} 
Moving towards symmetry breaking pattern, first the $\rm A_4$ flavor symmetry is broken by the flavon fields. Then, the spontaneous breaking of $\rm {B-L}$ gauge symmetry is implemented by assigning non-zero VEV to the scalar singlets $\phi_2$, $\phi_4$ and $\phi_8$. Finally, the Higgs doublet breaks the SM gauge symmetry to a low energy theory. The VEV alignment of the flavons are chosen for a specific case, where the solution of such VEV structures are described in various literature  \cite{Kobayashi:2008ih,King:2011zj,Altarelli:2005yp,Feruglio:2009iu,Borah:2017qdu} and are denoted as follows 
\bea
&&\langle H \rangle = \frac{v}{\sqrt{2}}\begin{pmatrix}0\\
1\end{pmatrix}, \;\; ~\langle \phi_2 \rangle = \frac{v_2}{\sqrt{2}}, \;\; ~ \langle \phi_4 \rangle = \frac{v_4}{\sqrt{2}},\;\; ~ \langle \phi_8 \rangle = \frac{v_8}{\sqrt{2}}, \nn\\
&&  \langle \phi_T \rangle = \frac{v_T}{\sqrt{2}} \begin{pmatrix}1\\
0\\0 \end{pmatrix},~\; \; \langle \chi \rangle = \frac{v_\chi}{\sqrt{2}} \begin{pmatrix}1\\
1\\1 \end{pmatrix}.\label{vev-scalar}
\eea
\subsection{CP-odd and CP-even scalar mass matrices}
The scalar  fields $H=(H^+,H^0)^T$ and $\phi_i$,  $(i=2,4,8)$ can be parametrized in terms of real scalars ($h_i$) and pseudo scalars ($A_i$) as
\begin{align}
&H^0 =\frac{1}{\sqrt{2} }(v+h)+  \frac{i}{\sqrt{2} } A^0\,, \nonumber \\
& \phi^0_i = \frac{1}{\sqrt{2} }(v_i+h_i)+  \frac{i}{\sqrt{2} } A_i\,,
\end{align}
where $v$ and $v_i$'s are the corresponding vacuum expectation values.
The CP-even component, $h$ of the scalar doublet $H$, is considered to be the observed Higgs boson at LHC with mass $M_h = 125$ GeV. We neglect the mixing of Higgs with the new CP-even scalars, $h_2,~h_4$ and $h_8$ (corresponding to $\phi_2,\phi_4$ and $\phi_8$). The mass matrix in the basis ($h_2, h_4,h_8$) takes the form
\begin{equation}
M_{E}^2=\begin{pmatrix}
 2 \lambda_{22} v_2^2 & v_2 \left(\sqrt{2}\mu_{24}+\lambda_{24} v_4+\lambda_{248}  v_8\right) & v_2 (\lambda_{248}  v_4+\lambda_{28}  v_8) \\
  v_2 \left(\sqrt{2}\mu_{24}+\lambda_{24} v_4+\lambda_{248}  v_8\right)~ &~ 2 \lambda_4 {v_4}^2-\frac{v_2^2 \left(\sqrt{2}\mu_{24} +\lambda_{248} v_8\right)}{2 v_4} ~&~ \frac{\lambda_{248}v_2^2}{2}+v_4 \left(\sqrt{2} \mu_{48} +\lambda_{48} v_8\right) \\
 v_2 \left(\lambda_{248} v_4+\lambda_{28} v_8\right) & \frac{\lambda_{248}v_2^2}{2}+v_4 \left(\sqrt{2} \mu_{48} +\lambda_{48} v_8\right) & 2 \lambda_8 {v_8}^2-\frac{ \left(\sqrt{2}\mu_{48}v^2_4 +\lambda_{248} v^2_2v_4\right)}{2 v_8} 
 \end{pmatrix}.
\end{equation}
The diagonalization of the above mass matrix results the mass eigenstates, represented by $H^\prime_{1}, H^\prime_{2}$ and $H^\prime_{3}$ with masses $M_{H^\prime_1}$, $M_{H^\prime_2}$, $M_{H^\prime_3}$ respectively. Moving to the CP-odd components, $A_2, A_4$ and $A_8$ (corresponding to $\phi_2,\phi_4$ and $\phi_8$), the mass matrix in the basis ($A_2, A_4,A_8$) is given by
\begin{equation}
M_{O}^2=\begin{pmatrix}
 -2 v_4 \left(\sqrt{2}\mu_{24} +\lambda_{248} v_8\right) & v_2 \left(\sqrt{2} \mu_{24}-\lambda_{248} v_8\right) & \lambda_{248} v_2 v_4 \\
 v_2 \left(\sqrt{2} \mu_{24}-\lambda_{248} v_8\right) & -\frac{\left(\sqrt{2}\mu_{24} +\lambda_{248} v_8\right) v_2^2}{2 v_4}-2 \sqrt{2}\mu_{48}  v_8 & \frac{\lambda_{248} v_2^2}{2}+\sqrt{2}\mu_{48}  v_4 \\
\lambda_{248} v_2 v_4 & \frac{\lambda_{248} v_2^2}{2}+\sqrt{2}\mu_{48}  v_4 & -\frac{v_4 \left(\lambda_{248} v_2^2+\sqrt{2} \mu_{48}  v_4\right)}{2 v_8} \\
\end{pmatrix}.
\end{equation}
The above mass matrix upon diagonalization, gives one massless eigenstate, to be absorbed by $U(1)$ boson, $Z^\prime$ and two massive modes (represented by $A^\prime_{1}$ and $A^\prime_{2}$ with masses $M_{A^\prime_1}$ and $M_{A^\prime_2}$ respectively),which remain as massive physical CP-odd scalars in the present framework. The gauge boson $Z^\prime$ attains the mass $M_{Z^\prime} = g_{BL} \left(4 v^2_2 + 16 v^2_4 + 64 v^2_8\right)^{1/2}$.
\subsection{Lagrangian and Leptonic Mass matrix}

The Yukawa interaction Lagrangian for charged leptons, allowed by the symmetries of the model is as follows
\begin{eqnarray}
\mathcal{L_\ell}&=& -H \left[\frac{y_e}{\Lambda} \left(\overline{L_L} {\phi_T}\right)_1 \otimes e_R +\frac{y_\mu}{\Lambda} \left(\overline{L_L} \phi_T \right)_{1'} \otimes \mu_R+\frac{y_\tau}{\Lambda} \left(\overline{L_L} \phi_T \right)_{1''} \otimes \tau_R \right]+{\rm ~H.c}.
\label{chargedlepton lagrangian}
\end{eqnarray}
The charged lepton mass matrix can be obtained from the above Lagrangian, which is written as follows,
\begin{eqnarray}
\mathcal{M}_\ell=\begin{pmatrix}
\frac{y_e v v_T}{2 \Lambda} && 0 && 0\\
0 && \frac{y_\mu v v_T}{2 \Lambda} && 0\\
0 && 0 && \frac{y_\tau v v_T}{2 \Lambda}\\
\end{pmatrix}.
\end{eqnarray}
Here, we use the VEV alignment of the flavon field $\phi_T$ as shown in Eqn. (\ref{vev-scalar}).  The ratio of expectation value of $\phi_T$ to the cut-off scale $\Lambda$ (i.e., $\frac{v_T}{\Lambda}$) is considered to be $\sim{\cal O}(0.1)$ for the typical scale $\Lambda\sim{\cal O}(10^9)$ GeV. The Yukawa couplings $y_e, y_\mu$ and $y_\tau$ are assumed to be hierarchical to get the appropriate masses for the charged leptons.

With the absence of  right-handed neutrinos ($\nu_R$) with $\rm B-L$ charge $-1$, there is no generic Dirac neutrino mass term i.e., $\bar{L} \tilde{H} \nu_R$ in the present framework. In addition, we cannot write  the Dirac and Majorana mass terms for light active neutrinos with the exotic fermions at tree level. Also the well known dimension-5 Weinberg operators ($\frac{LLHH}{\Lambda}$)  are not allowed, as these terms are not invariant under $\rm U(1)_{B-L}$ gauge symmetry. Therefore, the  neutral lepton masses are generated by the dimension six and seven operators. We thus obtain the Majorana mass terms for the neutrinos as 
\begin{eqnarray}
\mathcal{L_\nu}&=&-\frac{y_1}{{\Lambda}^2}[LLHH]_1 \otimes \phi_2-\frac{y_\chi}{\Lambda^3}[LLHH]_3\otimes\chi\otimes \phi_2.
\label{5dim mass term}
\end{eqnarray}
After the spontaneous symmetry breaking, the $3\times 3$ mass matrix for active neutrinos takes the form
\begin{eqnarray}
 \mathcal{M_\nu}=\begin{pmatrix}
a+\frac{2d}{3} &&  -\frac{d}{3} && -\frac{d}{3}\\
-\frac{d}{3} && \frac{2d}{3} && a-\frac{d}{3}\\
-\frac{d}{3} && a-\frac{d}{3} && \frac{2d}{3}
\end{pmatrix},
\end{eqnarray}
where,  $a=\frac{y_1 v_2 v^2}{2 \sqrt{2} \Lambda^2}$ and $d=\frac{y_\chi v_\chi v^2 v_2}{4 \Lambda^3}$. The sub-eV scale light neutrinos  can be obtained from the representative set of input model parameters as follows:  $v=246$ GeV, $\frac{v_\chi}{\Lambda} \approx 0.2$, $v_2 \approx \mathcal{O}(10^3)$ GeV and $y_1\approx y_\chi \approx \mathcal{O}(1)$. The flavor structure of the matrix ${\cal M}_\nu$ obtained from $\rm A_4$ symmetry is known to be diagonalized by the TBM mixing matrix \cite{Ma:2009wi}, which is given by
\begin{eqnarray}
 U_{\rm TBM}=\begin{pmatrix}
 -\sqrt{\frac{2}{3}} &\sqrt{\frac{1}{3}} & 0\\
 \sqrt{\frac{1}{6}}  & \sqrt{\frac{1}{3}} &-\sqrt{\frac{1}{2}}\\
 \sqrt{\frac{1}{6}}  & \sqrt{\frac{1}{3}} & \sqrt{\frac{1}{2}}\\
 \end{pmatrix}.
 \end{eqnarray}
 The standard scenario of three neutrinos gives a TBM mixing pattern  with vanishing reactor mixing angle in the framework of $\rm A_4$ flavor symmetry, which has been studied in various works in the  literature \cite{Memenga:2013vc,Meloni:2017cig,Channey:2018cfj,Garg:2017mjk,Dev:2016bml,Dev:2010pf}. In the next section we consider the active-sterile neutrino mixing by introducing an eV scale sterile-like neutrino and discuss the diagonalization of $4\times 4$ mass matrix to account for nonzero reactor mixing angle, consistent with current neutrino oscillation data.
\section{Neutrino masses and mixing with one eV scale sterile-like neutrino}
The standard scenario of three neutrino species has already been widely discussed in the literature, but the current experimental discrepancies from MiniBooNE and LSND data hint towards the possible existence  of the fourth neutrino. From the nomenclature of the sterile neutrinos, one can infer that it doesn't interact with the SM particles directly as it does not have gauge interaction, instead mixes with the active neutrinos during oscillation. The mixing between the flavor ($\nu_f$) and mass eigenstates ($\nu_i$) are related by
\begin{eqnarray}
\nu_f = \sum^n_{i=1} U_{i} \nu_i, \, .
\end{eqnarray}
where $n$ denotes the number of neutrino species. By considering three generations of active neutrinos, along with $n_s$ number of massive sterile species, one can have $n=3+n_s$ dimensional neutrino mixing matrix. In general, the mixing matrix will have $n-1 = n_s + 2$ Majorana phases, $3 \times (n-2) = 3 \times (n_s+1)$ mixing angles and $2n-5=2 n_s+1$ Dirac phases. Hence, in one sterile neutrino scenario, we will have $6$ mixing angles, $3$ Dirac phases and $3$ Majorana phases. The standard parameterization for $4\times 4$ neutrino mixing is given by
\begin{equation}
U = R_{34}\tilde{R}_{24}\tilde{R}_{14}R_{23}\tilde{R}_{13}R_{12}P \,
, \label{eq:UPMNS4}
\end{equation}
where the matrices $R_{ij}$ are rotations in $ij$ space and have the form
\begin{equation}
R_{34} = \begin{pmatrix} 1 & 0 & 0 & 0 \\ 0 & 1 & 0 & 0 \\ 0 & 0 &
c_{34} & s_{34} \\ 0 & 0 & -s_{34} & c_{34} \end{pmatrix}, ~\tilde{R}_{14} = \begin{pmatrix} c_{14} & 0 & 0 &
s_{14}e^{-i\delta_{14}} \\ 0 & 1 & 0 & 0 \\ 0 & 0 & 1 & 0 \\
-s_{14}e^{i\delta_{14}} & 0 & 0 & c_{14} \end{pmatrix}, ~\tilde{R}_{24} = \begin{pmatrix} 1 & 0 & 0 & 0 \\ c_{24} & 0 & 0 &
s_{24}e^{-i\delta_{24}} \\  0 & 0 & 1 & 0 \\
-s_{24}e^{i\delta_{24}} & 0 & 0 & c_{24} \end{pmatrix}. \\
\end{equation}
Here, $s_{ij} = \sin\theta_{ij}$, $c_{ij}=\cos\theta_{ij}$ and $P$ denotes the diagonal matrix with three Majorana phases $\alpha$,
$\beta$ and $\gamma$,
\begin{equation} P = {\rm
diag}\left(1,e^{i\alpha/2},e^{i(\beta/2+\delta_{13})},e^{i(\gamma/2+\delta_{14})}\right)\, .
\end{equation}
 The current experimental searches of light sterile neutrino prefer a larger value of active-sterile mass squared differences than the observed solar and atmospheric mass squared differences of active neutrino oscillation. This implies that the mass for sterile neutrino can be either heavier or lighter than the active ones.  We know that in normal ordering the active neutrinos have the form $m_3 \gg m_2 \textgreater m_1$ whereas the inverse ordering is given by $m_2 \textgreater m_1 \gg m_3$). So accordingly, there will be four possibilities in mass orderings if a sterile neutrino is added to the framework. Following the top to bottom nomenclature, i.e. if $m_s \gg m_{1,2,3}$, can be denoted as 1+3 scenario for  normal or inverted ordering of the active neutrinos. Whereas, if the case is reversed, i.e. if sterile state is lighter than the active ones ($m_{1,2,3}\gg m_s$), this configuration is named as 3+1 model in literature. Moreover, less attention is given to 3+1 scenario as they are prone to  conflict with several experimental observations. In this model, we  consider 1+3 like scenarios, to explain the neutrino phenomenology with eV scale exotic fermion. 

\begin{table}[t]
 \centering
 \caption{ $2\sigma$ estimated values of the mixing parameters in one sterile neutrino scenario ~\cite{Kopp:2011qd}.}
 \label{table:osc_params}
 \vspace{2mm}
 \begin{tabular}{lccccc}
\hline \hline
& parameter & $\Delta m^2_{41}$ [eV] & $|U_{e4}|^2$ \\ 
\hline
 \multirow{2}{*}{3+1/1+3} & best-fit & 1.78 & 0.023 & & \\
 & $2\sigma$ & 1.61--2.01 & 0.006--0.040  \\
  \hline
\end{tabular}
\end{table}
Table~\ref{table:osc_params} shows the best-fit and $2\sigma$ ranges of the relevant oscillation  parameters, we used for this work.
\subsection{Diagonalization of Neutrino Mass matrix in $3+1$ like scenario}
Out of the three exotic fermions, we consider the fermion with $\rm B-L$ charge 5 to mix with the SM neutrinos, to study the neutrino phenomenology analogous to the mixing between the standard and sterile neutrinos $3+1$ mixing scenario. The Majorana mass terms for fermions and their interaction with SM leptons are given by
\begin{eqnarray}
L_{N}=-\Big[\frac{y_{11}}{\Lambda} \left (N_1 N_1\phi_8^\dagger \phi_2^\dagger\right) +\frac{y_s}{\Lambda^3}\left (\overline{L}\tilde{H} N_{1}(\phi^\dagger_4 \phi^\dagger_2+\phi^\dagger_8\phi_2)\chi \right)+ {\rm H.c.}\Big].
\label{sterile lag}
\end{eqnarray}
Here, the first and second terms are for eV scale Majorana mass for sterile-like neutrino, generated from dimension 5 and 7 operators respectively. The dimension 7 operator in the third term induces active-sterile neutrino mixing. Thus the resulting $4\times 4$ neutrino mass matrix is given as follows\\
\begin{equation}
\mathcal{M_\nu}=\begin{pmatrix}
a+\frac{2d}{3} &&  -\frac{d}{3} && -\frac{d}{3} && e\\
-\frac{d}{3} && \frac{2d}{3} && a-\frac{d}{3} && e\\
-\frac{d}{3} && a-\frac{d}{3} && \frac{2d}{3} && e\\
e && e && e && m_s
\end{pmatrix}.
\label{mass matrix}
\end{equation}
Here, $e=\frac{y_s v (v_4+v_8) v_2 v_\chi}{4 \Lambda^3}$ and $m_s\approx\frac{y_{11} v_2 v_8}{2 \Lambda}$ (neglecting the contributions from dim-6 operators), are related to the active-sterile mixing parameter $(e/m_s)$ and Majorana mass of the sterile-like fermion respectively. The eV scale mass of the sterile neutrino ($m_s$) and sub-eV scale parameter, responsible  for the active-sterile neutrino mixing 
($e$) can be generated with the following representative set of input model parameters: $y_{11}\approx \mathcal{O}(10^{-6})$,  $y_s\approx\mathcal{O}(1)$, $v_2=v_4=v_8 \simeq \mathcal{O}(10^3)$ GeV and $\frac{v_\chi}{\Lambda}\approx 0.2$. The above mass matrix is analytically diagonalized to get the physical masses of $3+1$ neutrinos and the eigenvector matrix is given as
\begin{equation}
  U = \begin{pmatrix} \frac{2}{\sqrt{6}} & \frac{1}{6e}\frac{K_-}{N_-} & 0 & \frac{1}{6e}\frac{K_+}{N_+} \\[2mm] -\frac{1}{\sqrt{6}} &  \frac{1}{6e}\frac{K_-}{N_-} & -\frac{1}{\sqrt{2}} & \frac{1}{6e}\frac{K_+}{N_+} \\[2mm] -\frac{1}{\sqrt{6}} & \frac{1}{6e}\frac{K_-}{N_-} & \frac{1}{\sqrt{2}} & \frac{1}{6e}\frac{K_+}{N_+} \\[2mm] 0 & \frac{1}{N_-} & 0 & \frac{1}{N_+} \end{pmatrix}\, ,
 \label{eq:v4_ex}
\end{equation}
where, $K_{\pm}=a-m_s\pm\sqrt{12e^2+(a-m_s)^2}$ and $N_{\pm}^2=1+\frac{\left(a-m_s \pm \sqrt{12 e^2+(a-m_s)^2}\right)^2}{12 e^2}$. 
If one assumes that $a < m_s$ and expands to second order in the small ratio $e/m_s$, the resulting mixing matrix is given by \cite{Barry:2011wb}
\begin{equation}
   U \simeq \begin{pmatrix} \frac{2}{\sqrt{6}} & \frac{1}{\sqrt{3}} & 0 & 0 \\ -\frac{1}{\sqrt{6}} & \frac{1}{\sqrt{3}} & -\frac{1}{\sqrt{2}} & 0 \\ -\frac{1}{\sqrt{6}} & \frac{1}{\sqrt{3}} & \frac{1}{\sqrt{2}} & 0 \\ 0 & 0 & 0 & 1 \end{pmatrix} + \begin{pmatrix} 0 & 0 & 0 & \frac{e}{m_s} \\ 0 & 0 & 0 & \frac{e}{m_s} \\ 0 & 0 & 0 & \frac{e}{m_s} \\ 0 & -\frac{\sqrt{3}e}{m_s} & 0 & 0 \end{pmatrix} + \begin{pmatrix} 0 & -\frac{\sqrt{3} e^2}{2m_s^2} & 0 & 0 \\ 0 & -\frac{\sqrt{3} e^2}{2m_s^2} & 0 & 0 \\ 0 & -\frac{\sqrt{3} e^2}{2 m_s^2} & 0 & 0 \\ 0 & 0 & 0 & -\frac{3e^2}{2 m_s^2}\end{pmatrix}.
\label{eq:v4}
\end{equation}
But this mixing pattern predicts $\theta_{13}=0$, which has already been experimentally ruled out. Hence, to explain the non-zero   $\theta_{13}$, we introduce one extra flavon field $\zeta$, which is charged as $1'$ under $A_4$ symmetry, in order to perturb the neutrino mass matrix from the TBM mixing pattern. Including this new flavon field, the perturbed Lagrangian is given by
\begin{equation}
\mathcal{L}_p= -\frac{y_p}{\Lambda^3} \left([LLHH]_{1^{''}}\otimes(\zeta)_{1'}\otimes\phi_2+{\rm H.c}\right).
 \end{equation}
 When the flavon field acquires VEV, $\langle \zeta \rangle =\frac{v_\zeta}{\sqrt{2}}$, the above term contributes to the mass matrix in \eqref{mass matrix}. Hence, the modified neutrino mass matrix can be written as
\begin{equation}
\mathcal{M_\nu}=\begin{pmatrix}
a+\frac{2d}{3} &&  -\frac{d}{3} && -\frac{d}{3} && e\\
-\frac{d}{3} && \frac{2d}{3} && a-\frac{d}{3} && e\\
-\frac{d}{3} && a-\frac{d}{3} && \frac{2d}{3} && e\\
e && e && e && m_s
\end{pmatrix}+\begin{pmatrix}
0 && 0 && b && 0\\
0 && b && 0  && 0\\
b && 0 && 0 && 0\\
0 && 0 && 0 && 0
\end{pmatrix},
\end{equation}
where, $b=\frac{y_p v^2 v_\zeta v_2}{4 \Lambda^3}$. We choose $\frac{v_\zeta}{\Lambda}\approx \mathcal{O}(0.1)$ and $y_p \approx \mathcal{O}(1)$ in order to achieve a sub-eV scale for the parameter $b$. We analytically diagonalize the above mass matrix and the mixing matrix is constructed from the normalized eigenvectors, which takes the form 
\begin{equation}
  U = \begin{pmatrix} \frac{-p_+}{l_{p+}} & \frac{1}{6e}\frac{K_{p-}}{N_{p-}} & \frac{-p_-}{l_{p-}} & \frac{1}{6e}\frac{K_{p+}}{N_{p+}} \\[2mm] \frac{q_+}{l_{p+}} &  \frac{1}{6e}\frac{K_{p-}}{N_{p-}} & \frac{q_-}{l_{p-}} & \frac{1}{6e}\frac{K_{p+}}{N_{p+}} \\[2mm] \frac{1}{l_{p+}} & \frac{1}{6e}\frac{K_{p-}}{N_{p-}} & \frac{1}{l_{p-}} & \frac{1}{6e}\frac{K_{p+}}{N_{p+}} \\[2mm] 0 & \frac{1}{N_{p-}} & 0 & \frac{1}{N_{p+}} \end{pmatrix},
 \label{eq:v4_ex}
\end{equation}
here, 
\begin{eqnarray}
&& K_{p \pm}=a+b-m_s\pm\sqrt{12e^2+(a+b-m_s)^2} ,\nn \\
&& N_{p \pm}^2=1+\frac{\left(a+b-m_s \pm \sqrt{12 e^2+(a+b-m_s)^2}\right)^2}{12 e^2} , \nn  \\
&& p_{\pm}=\frac{a\pm \sqrt{a^2-ab+b^2}}{a-b}, \hspace{3mm}  q_{\pm}=\frac{b \pm \sqrt{a^2-ab+b^2}}{a-b} , \nn \\ 
&& l_{p \pm}^2=1+(p_\pm)^2+(q_\pm)^2.
\end{eqnarray}
And the mass eigenvalues of the $4\times4$ neutrino mixing matrix are stated as following
\begin{eqnarray}
&& m_{\nu_1}=d+\sqrt{a^2-ab+b^2}, \nn \\
&& m_{\nu_2}=\frac{1}{2}[a+b+m_s-\sqrt{12 e^2+(a+b-m_s)^2}], \nn \\
&& m_{\nu_3}=d-\sqrt{a^2-ab+b^2}, \nn \\
&& m_{\nu_4}=\frac{1}{2}[a+b+m_s+\sqrt{12 e^2+(a+b-m_s)^2}].
\label{masses}
\end{eqnarray} 
Comparing with the standard $4\times 4$ mixing matrix, we can have the mixing angles as follows\\
\begin{eqnarray}
&& \sin^2 \theta_{12} = \frac{|U_{e2}|^2}{1-|U_{e4}|^2} \simeq \frac{1}{3}\left[1 - 2\left(\frac{e}{m_s}\right)^2\right],\\
&& \sin^2 \theta_{23}= \frac{|U_{\mu3}|^2(1-|U_{e4}|^2)}{1-|U_{e4}|^2-|U_{\mu4}|^2} \simeq \frac{1}{2}\left[1 + \left(\frac{e}{m_s}\right)^2\right],\\
&& \sin^2 \theta_{14}= |U_{e4}|^2\approx \left(\frac{e}{m_s}\right)^2, \\
&& \sin \theta_{34} = \frac{|U_{\tau4}|^2}{1-|U_{e4}|^2-|U_{\mu4}|^2}\approx \left(\frac{e}{m_s}\right)^2,\\
&& \sin^2 \theta_{24}=\frac{|U_{\mu 4}|^2}{1-|U_{e4}|^2} \approx \left(\frac{e}{m_s}\right)^2,\\
&& \sin^2 \theta_{13}= \frac{|U_{e3}|^2}{(1-\sin^2{\theta_{23}})(1-\sin^2{\theta_{14}})}=\frac{b^2}{4 a^2}\left(1+2 \frac{e^2}{m^2_s}\right).
\end{eqnarray}
From the above equations, we can infer that adding the perturbation term in the interaction Lagrangian gives non-zero $\theta_{13}$, which is compatible with the current oscillation data.
\begin{figure}[t!]
\begin{center}
\includegraphics[width=51mm,height=46mm]{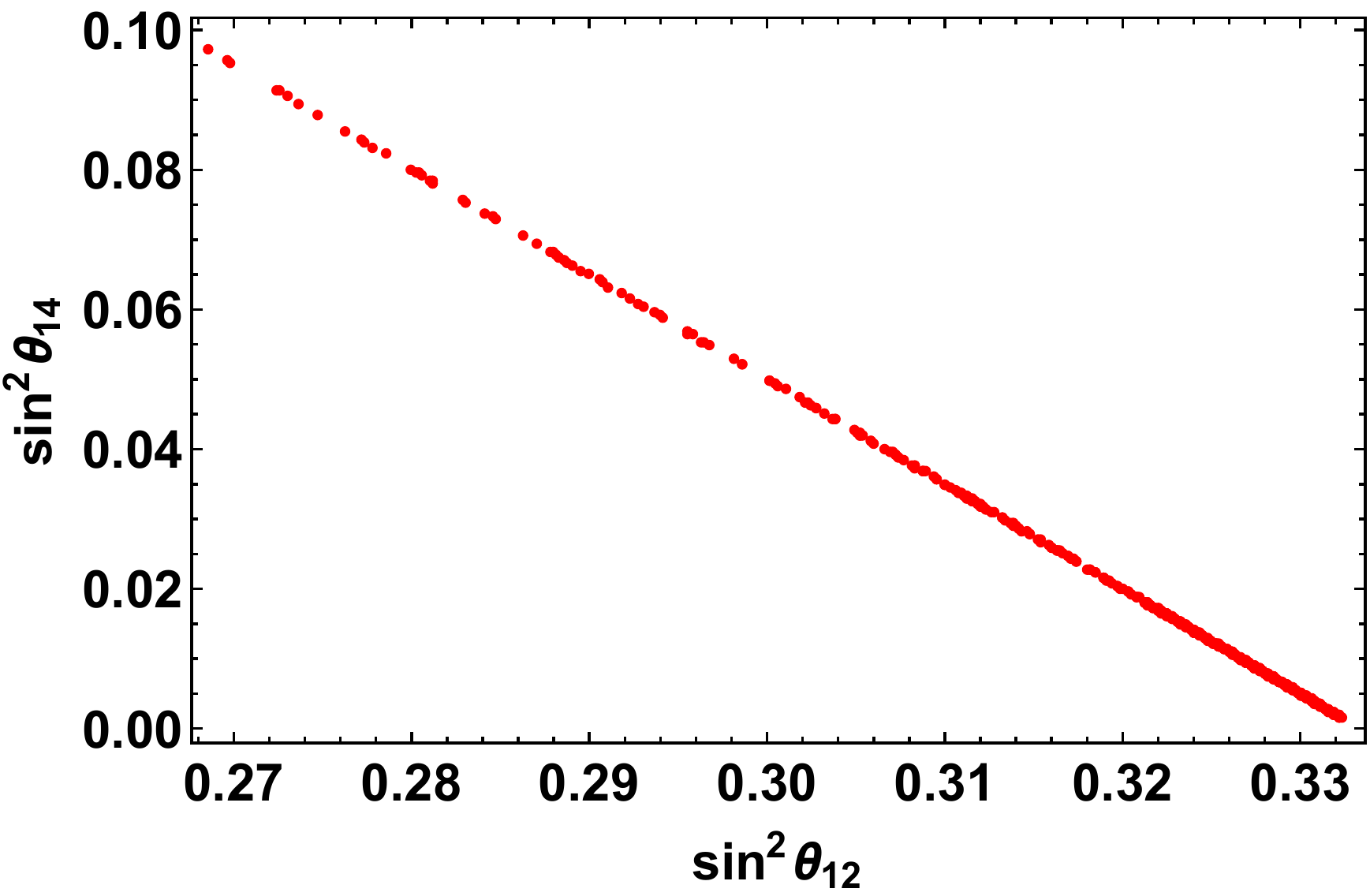}
\hspace*{0.2 true cm}
\includegraphics[width=51mm,height=46mm]{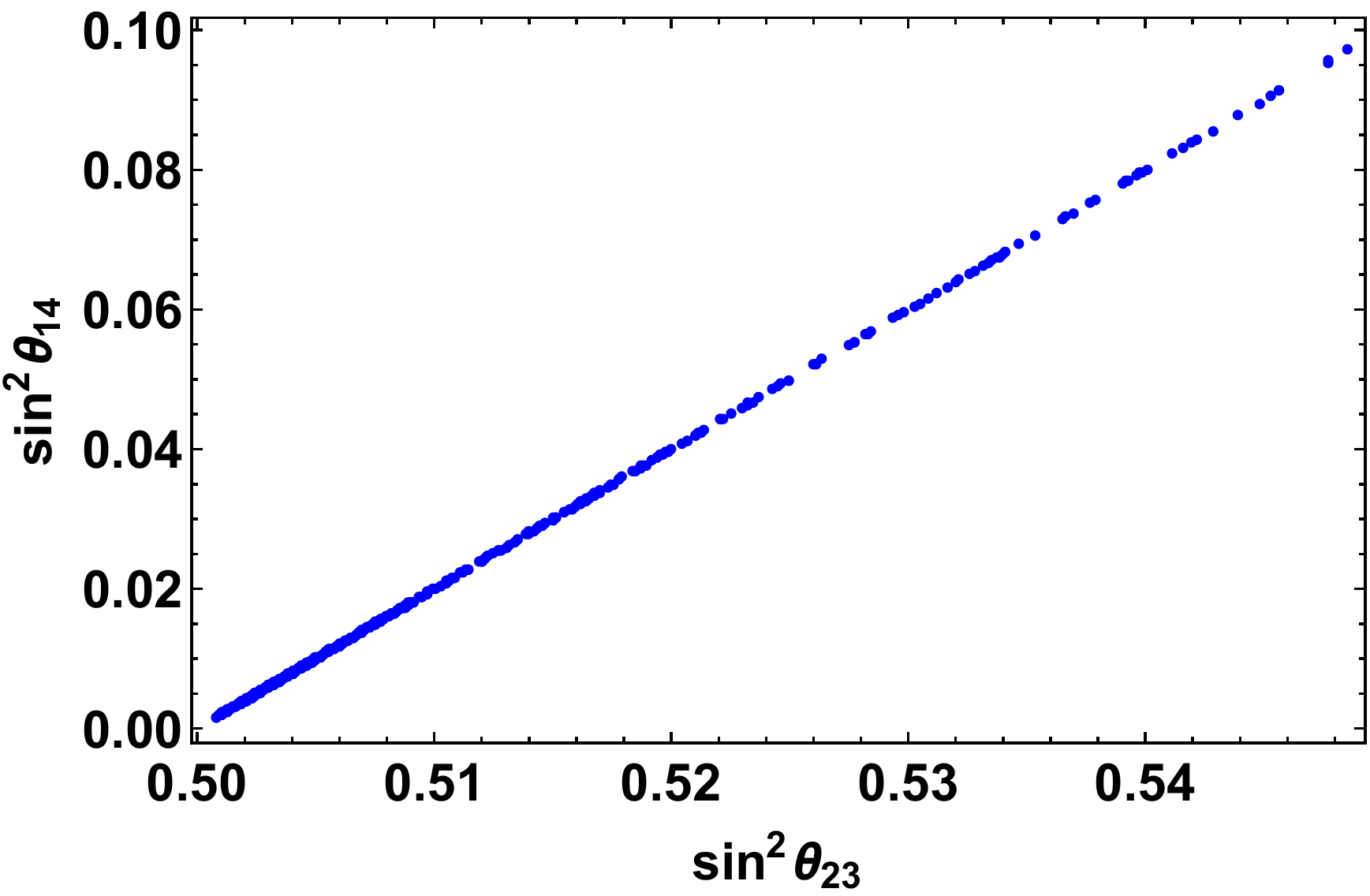}
\hspace*{0.2 true cm}
\includegraphics[width=51mm,height=46mm]{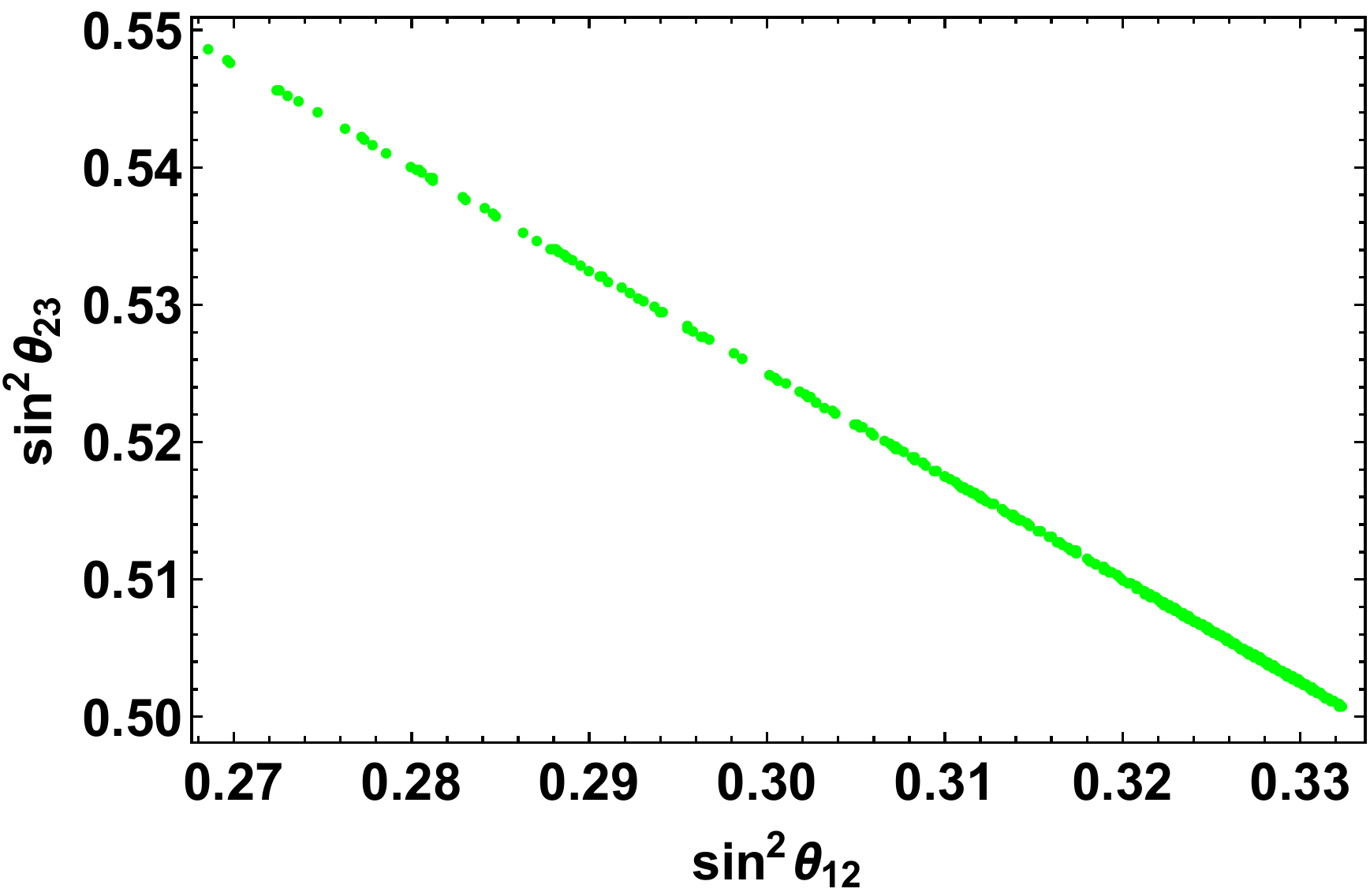}
\caption{Correlation of the active-sterile mixing angle $\theta_{14}$ with solar mixing  angle $\theta_{12}$ (left panel),  with atmospheric mixing angle $\theta_{23}$ (middle panel) and between the  atmospheric and solar mixing angles  (right panel).\label{fig1}}
\end{center}
\end{figure}
%

\subsection{Numerical Analysis}
To perform numerical analysis in a systematic way, we define  $\lambda_1=\frac{b}{a}$, $\lambda_2=\frac{d}{a}$ and $\lambda_3=\frac{e^2}{m_s a}$ with $\phi_{ba}$, $\phi_{da}$, $\phi_{ea}$ as the phases of $\lambda_1$, $\lambda_2$ and $\lambda_3$ respectively. The expressions of mass eigenvalues in \eqref{masses} can thus be written as
\begin{eqnarray}
&& m_{\nu_1}= |m_{\nu_1}|e^{i \phi_1}=|a| \Big |\lambda_2 e^{i \phi_{da}}+\sqrt{1-\lambda_1 e^{i \phi_{ba}}+{\lambda_1}^2 e^{2 i \phi_{ba}}} \Big | e^{i \phi_1}, \nn \\
&& m_{\nu_2}= |m_{\nu_2}|e^{i \phi_2}=|a| \Big |1+\lambda_1 e^{i \phi_{ba}}-3\lambda_3 e^{i \phi_{ea}}\Big |  e^{i \phi_2}, \nn \\
&& m_{\nu_3}= |m_{\nu_3}|e^{i \phi_3}=|a| \Big |\lambda_2 e^{i \phi_{da}}-\sqrt{1-\lambda_1 e^{i \phi_{ba}}+{\lambda_1}^2 e^{2 i \phi_{ba}}}\Big | e^{i \phi_3}\;, \nn \\
&& m_{\nu_4}= |m_{\nu_4}|e^{i \phi_4}=|a|  \Big |\frac{m_s}{a}+3\lambda_3 e^{i \phi_{ea}} \Big |e^{i \phi_4} .
\label{masses}
\end{eqnarray}
Thus, one obtains the physical masses as
\begin{eqnarray}
&& |m_{\nu_1}|=|a| \Big [(\lambda_2 \cos{\phi_{da}}+C)^2 + (\lambda_2 \sin{\phi_{da}}+D)^2\Big ]^{\frac{1}{2}}, \nn \\
&& |m_{\nu_2}|=|a| \Big[(1+\lambda_1 \cos{\phi_{ba}}-3\lambda_3 \cos{\phi_{ea}})^2+(\lambda_1 \sin{\phi_{ba}}-3\lambda_3 \sin{\phi_{ea}})^2\Big ]^{\frac{1}{2}}, \nn \\
&& |m_{\nu_3}|=|a| \Big [(\lambda_2 \cos{\phi_{da}}-C)^2 + (\lambda_2 \sin{\phi_{da}}-D)^2 \Big]^{\frac{1}{2}}, \nn \\
&& |m_{\nu_4}|=|a|\Big [(\frac{m_s}{a}+3\lambda_3 \cos{\phi_{ea}})^2+(3\lambda_3 \sin{\phi_{ea}})^2 \Big ]^{\frac{1}{2}}, 
\end{eqnarray}
where, 
\begin{eqnarray}
&& C=\left( \frac{A+\sqrt{A^2+B^2}}{2}\right)^\frac{1}{2}, \hspace{3mm}  D=\left( \frac{-A+\sqrt{A^2+B^2}}{2}\right)^\frac{1}{2}, \nn \\
&& A=1-\lambda_1 \cos{\phi_{ba}}+\lambda_1^2 \cos{2 \phi_{ba}}, \hspace{2mm} B=-\lambda_1 \sin{\phi_{ba}}+\lambda_1^2 \sin{2 \phi_{ba}}\;.
\end{eqnarray}
The corresponding phases in the mass eigenvalues values are given by
\begin{eqnarray}
&& \phi_1=\tan^{-1}\left[\frac{\lambda_2 \sin{\phi_{da}}+D}{\lambda_2 \cos{\phi_{da}}+C}\right], \nn \\
&& \phi_3=\tan^{-1}\left[\frac{\lambda_2 \sin{\phi_{da}}-D}{\lambda_2 \cos{\phi_{da}}-C}\right], \nn \\
&& \phi_2=\tan^{-1}\left[\frac{\lambda_1 \sin{\phi_{ba}}-3\lambda_3 \sin{\phi_{ea}}}{1+\lambda_1 \cos{\phi_{ba}}-3\lambda_3 \cos{\phi_{ea}}}\right], \nn \\
&& \phi_4=\tan^{-1}\left[\frac{3\lambda_3 \sin{\phi_{ea}}}{\frac{m_s}{a}+3\lambda_3 \cos{\phi_{ea}}}\right]. 
\end{eqnarray}
The model also predicts a  large CP violating Dirac phase, associated with the non-zero reactor mixing angle,  which can be obtained from \eqref{eq:UPMNS4}
\begin{equation}
{\rm Exp}(-i \delta_{13})= \frac{U_{13}}{\sin{\theta_{13}} \cos{\theta_{14}}\cos{\theta_{23}}} \approx   {\rm Exp}({i \phi_{ba}}).
\label{delta_cp}
\end{equation}
The above equation gives   $\sin{\delta_{13}} \approx -\sin{\phi_{ba}}$. 
To constrain the model parameters, compatible with the   $3\sigma$ limits of  the current oscillation data, we perform a random scan of these parameters over the following ranges:
\bea
&&a \in [-0.06,0.06]~ {\rm eV},~~~e\in [-0.1,0.1]~ {\rm eV},~~~~m_s \in [-1.5,1.5]~ {\rm eV},~~~~\lambda_1 \in[0.01,0.3],\nn\\
&& \lambda_2 \in [0.01,0.5]\;,~~~~~\phi_{ba,da,ea}\in[-\pi,\pi]\;,
\eea
and show the correlation plots between  different mixing angles in Fig. \ref{fig1}.
We now proceed to discuss explicitly the constraints on different parameters from the availed neutrino oscillation data for vanishing and non-vanishing Dirac CP phase, by fixing various model parameters. Firstly we filter the parameters from the constraints of observed solar and atmospheric mass squared differences in $3\sigma$ range and then the obtained parameter space is further constrained from the cosmological observation of total neutrino mass.
\begin{table}[htbp]
\centering
{
\renewcommand{\arraystretch}{1.2}
\catcode`?=\active \def?{\hphantom{0}}
\begin{minipage}{\linewidth}
\begin{tabular}{|l|c|c|c|}
\hline
Parameter & Best fit $\pm$ $1\sigma$ &  2$\sigma$ range& 3$\sigma$ range
\\
\hline\hline
$\Delta m^2_{21}\: [10^{-5}\eVq]$ & 7.56$\pm$0.19  & 7.20--7.95 & 7.05--8.14 \\
\hline
$|\Delta m^2_{31}|\: [10^{-3}\eVq]$ (NO) &  2.55$\pm$0.04 &  2.47--2.63 &  2.43--2.67\\
$|\Delta m^2_{31}|\: [10^{-3}\eVq]$ (IO)&  2.47$^{+0.04}_{-0.05}$ &  2.39--2.55 &  2.34--2.59 \\
\hline
$\sin^2\theta_{12} / 10^{-1}$ & 3.21$^{+0.18}_{-0.16}$ & 2.89--3.59 & 2.73--3.79\\
\hline
  $\sin^2\theta_{23} / 10^{-1}$ (NO)
	  &	4.30$^{+0.20}_{-0.18}$ 
	& 3.98--4.78 \& 5.60--6.17 & 3.84--6.35 \\
  $\sin^2\theta_{23} / 10^{-1}$ (IO)
	  & 5.98$^{+0.17}_{-0.15}$ 
	& 4.09--4.42 \& 5.61--6.27 & 3.89--4.88 \& 5.22--6.41 \\
\hline 
$\sin^2\theta_{13} / 10^{-2}$ (NO) & 2.155$^{+0.090}_{-0.075}$ &  1.98--2.31 & 1.89--2.39 \\
$\sin^2\theta_{13} / 10^{-2}$ (IO) & 2.155$^{+0.076}_{-0.092}$ & 1.98--2.31 & 1.90--2.39 \\
    \hline
  \end{tabular}
  \caption{ \label{tab:sum-2017} 
   The experimental values of Neutrino oscillation parameters for $1\sigma$, $2\sigma$ and $3\sigma$ range ~\cite{deSalas:2017kay,Gariazzo:2018pei}.}
    \end{minipage}
  }
\end{table}

\subsubsection{Variation of model parameters by fixing $\lambda_2=0.5$ and $\phi_{ba}=0$}
\begin{figure}[t!]
\begin{center}
\includegraphics[width=70mm,height=50mm]{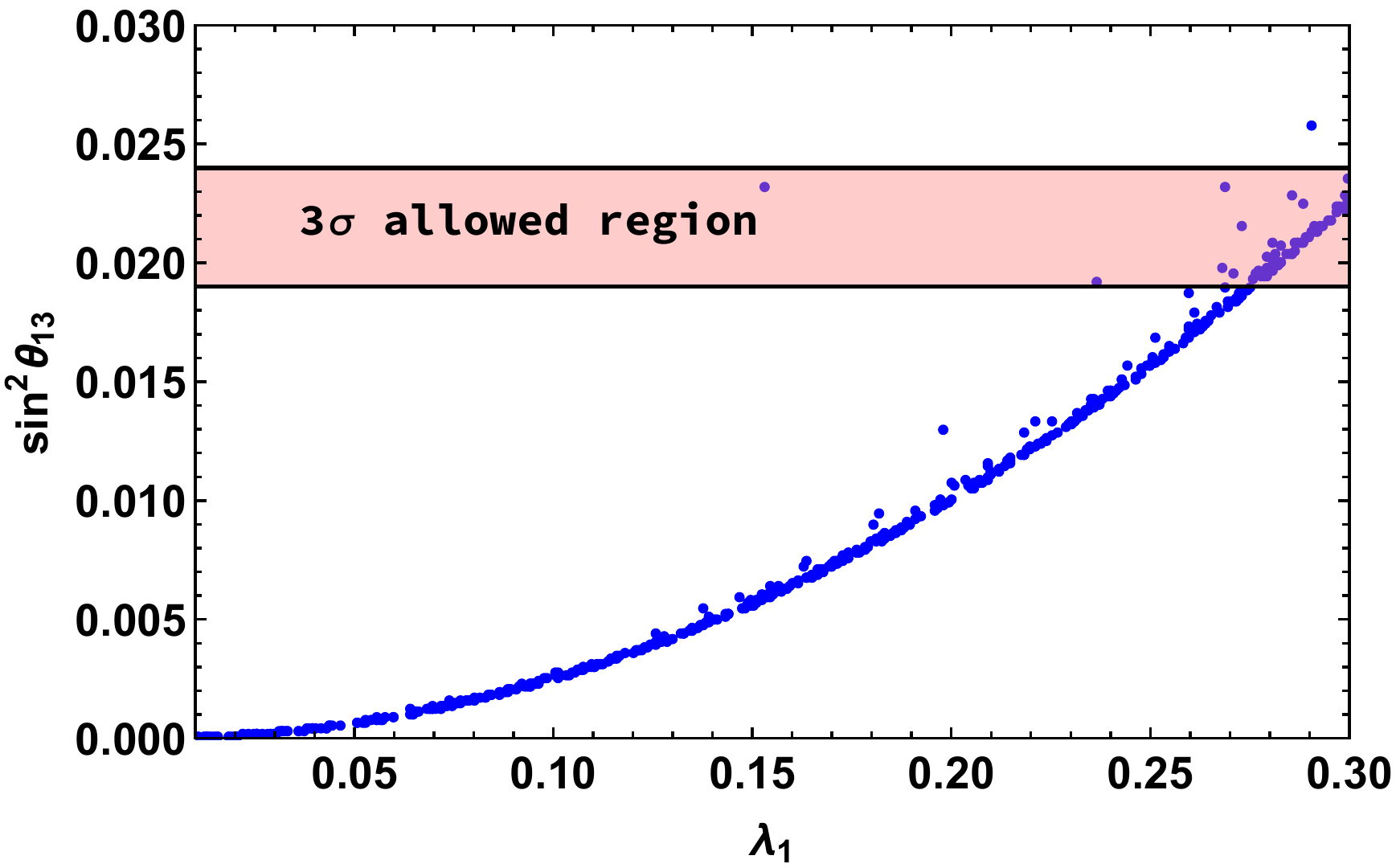}
\includegraphics[width=70mm,height=50mm]{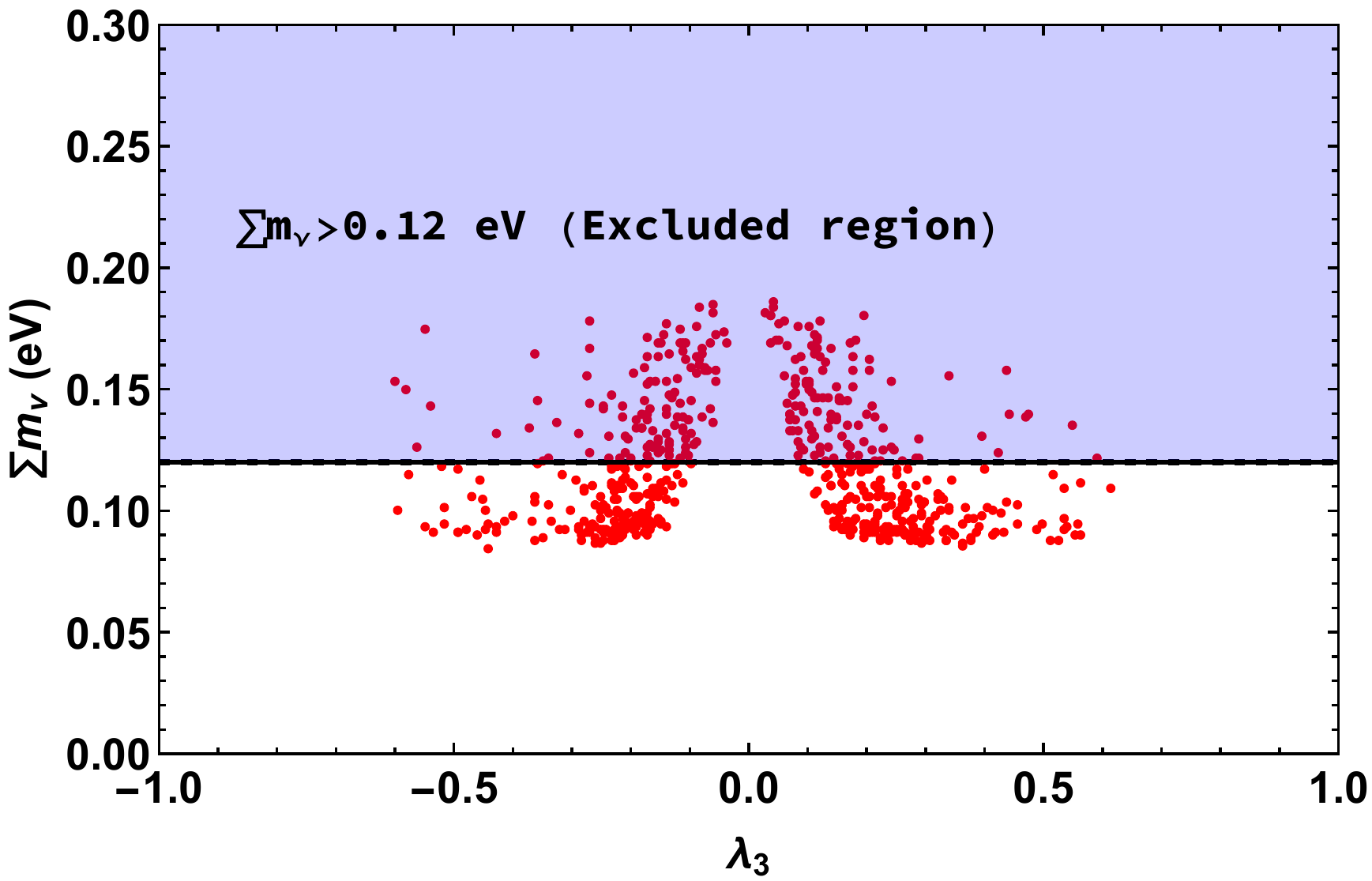}
\caption{The left panel shows the dependence of the reactor mixing angle $\theta_{13}$   on $\lambda_1$ and right panel represents the correlation between  the sum of total active neutrino masses and $\lambda_3$.}
\label{l1_s13_l3mt}
\end{center} 
\end{figure}

\begin{figure}[t!]
\begin{center}
\includegraphics[width=70mm,height=50mm]{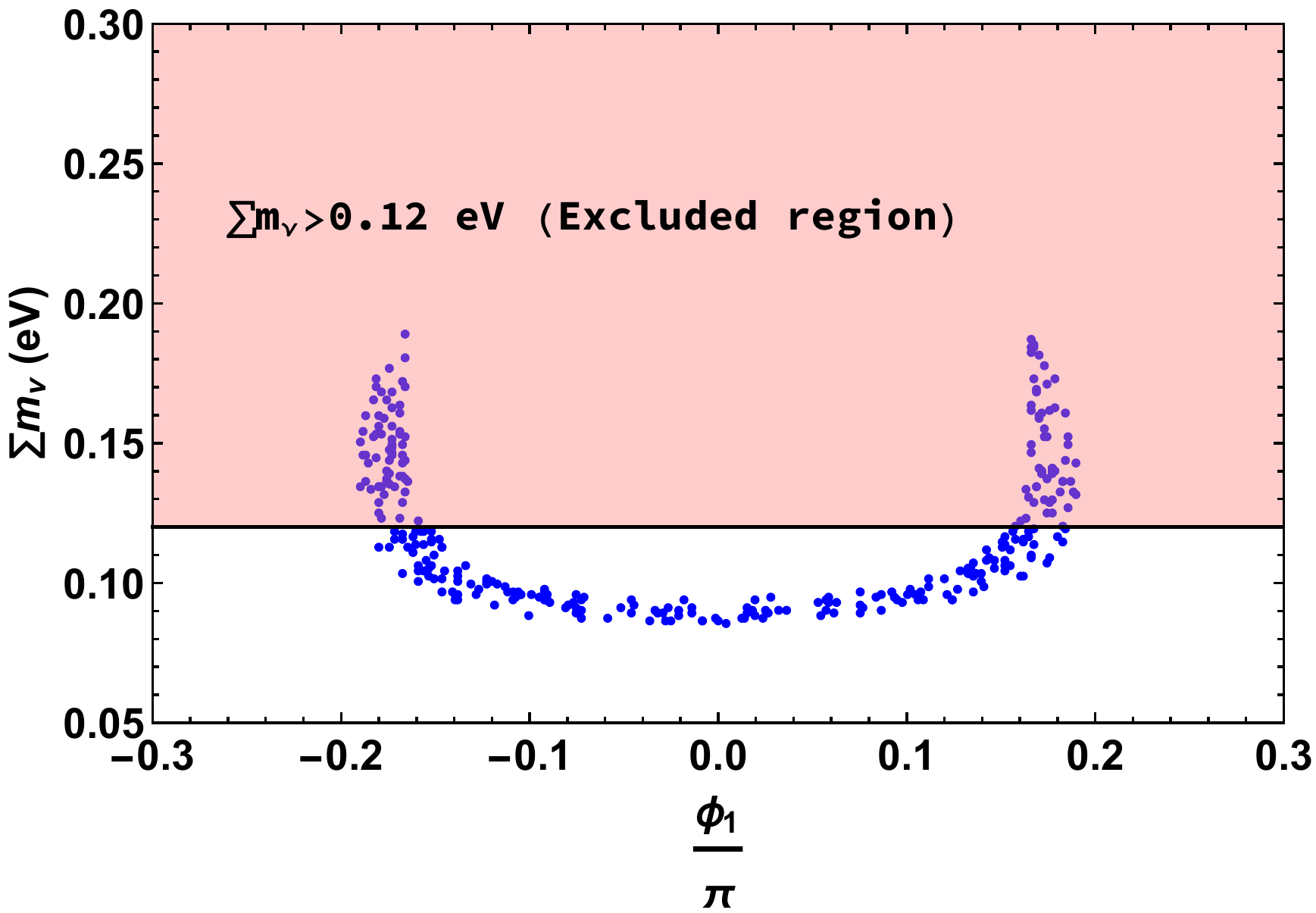}
\hspace*{0.1 true cm}
\includegraphics[width=70mm,height=50mm]{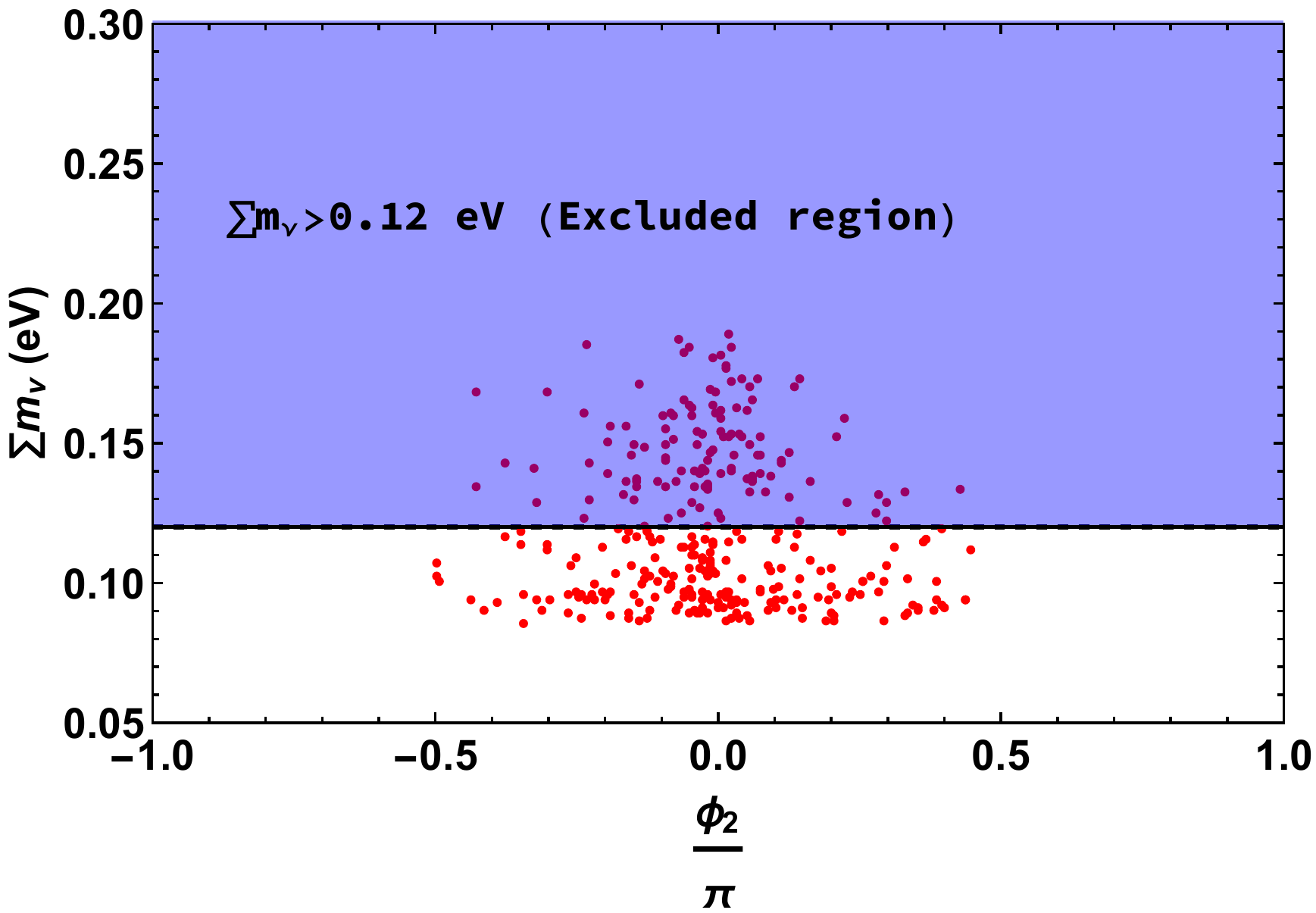}
\includegraphics[width=70mm,height=50mm]{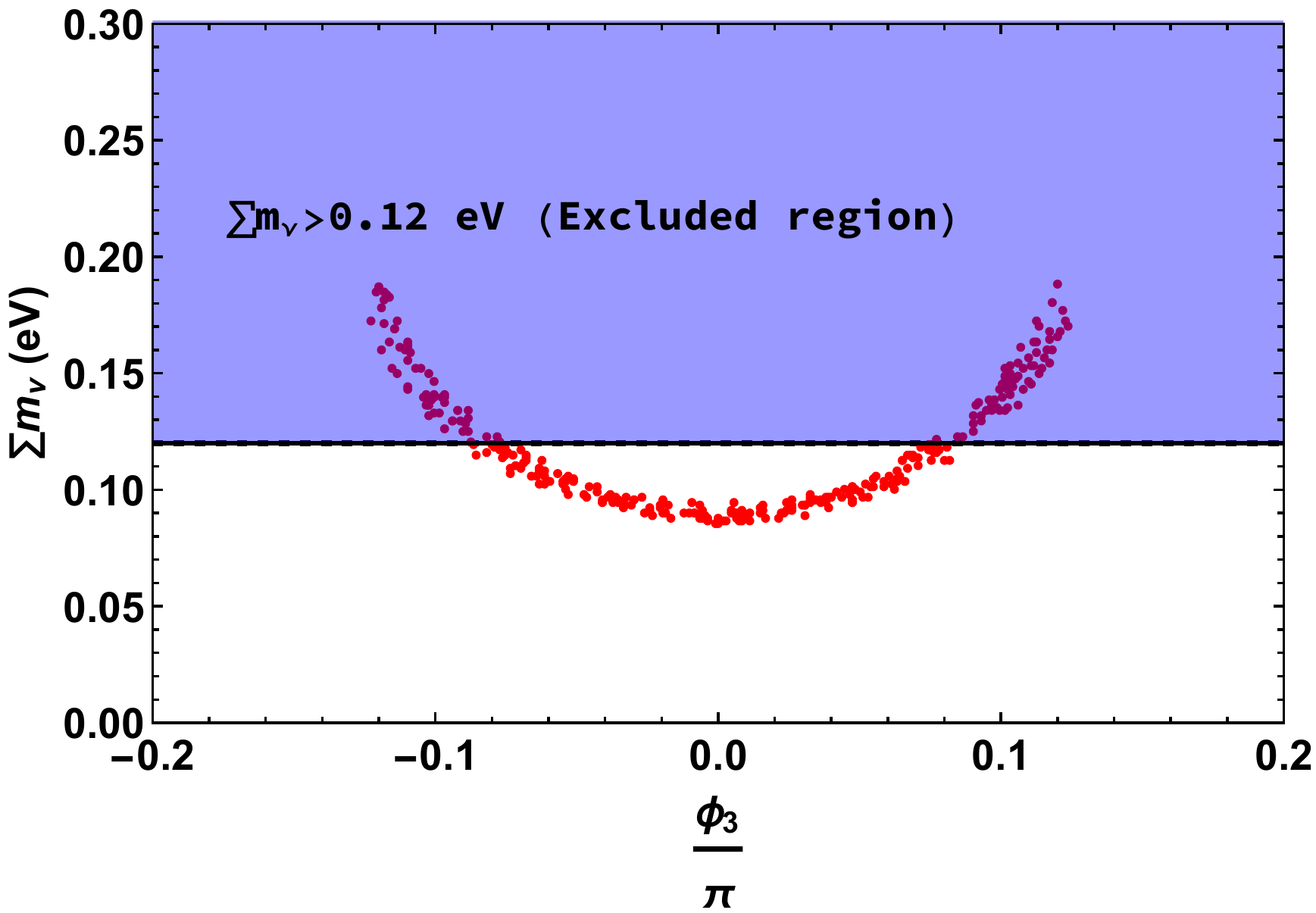}
\caption{Correlation between  $\phi_1$ (top-left panel), $\phi_2$  (top-right panel) and $\phi_3$ (bottom) with the total active neutrino mass.}
\label{p1p3_mt}
\end{center}
\end{figure}

\begin{figure}
\begin{center}
\includegraphics[width=70mm,height=50mm]{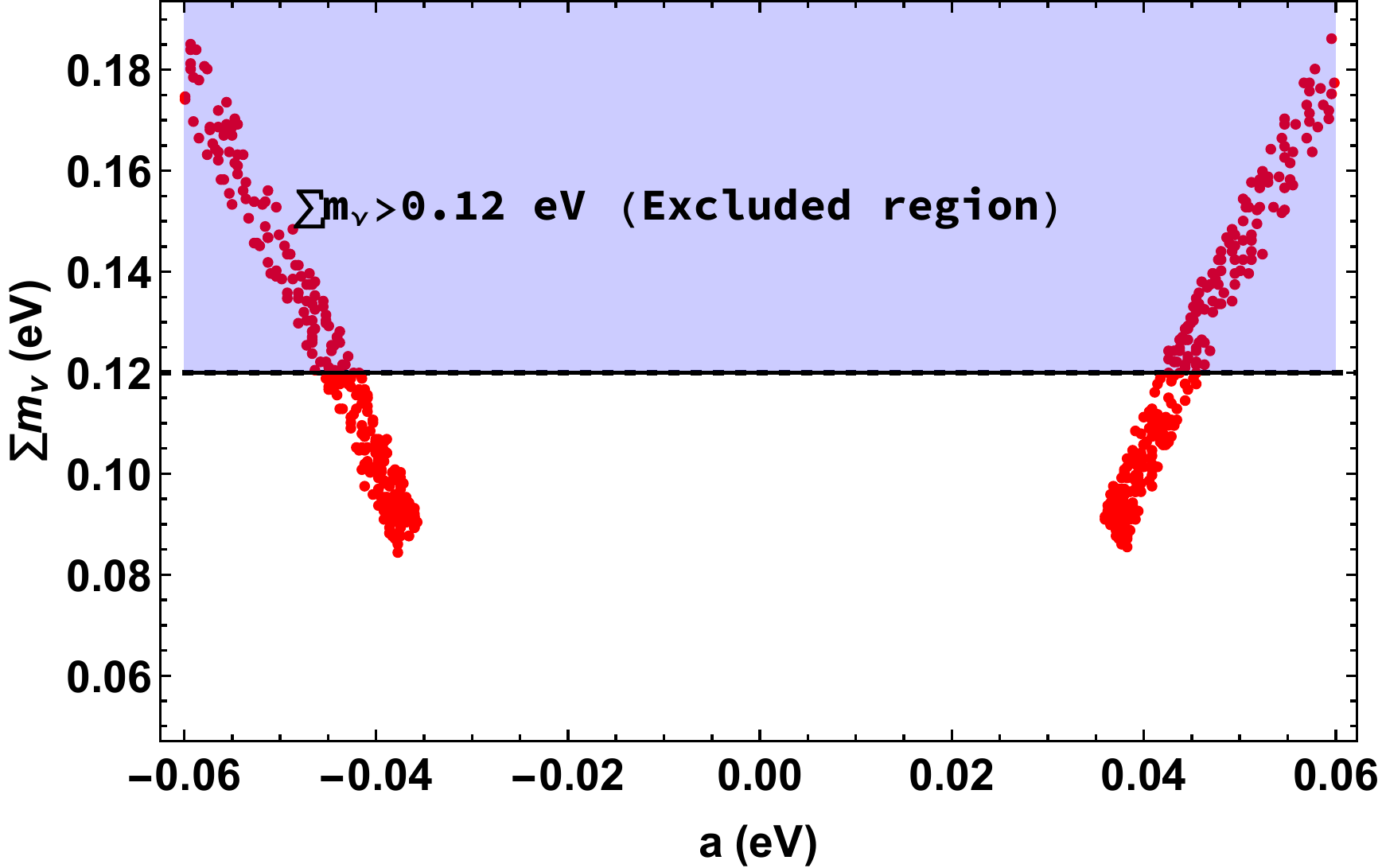}
\hspace*{0.1 true cm}
\includegraphics[width=70mm,height=50mm]{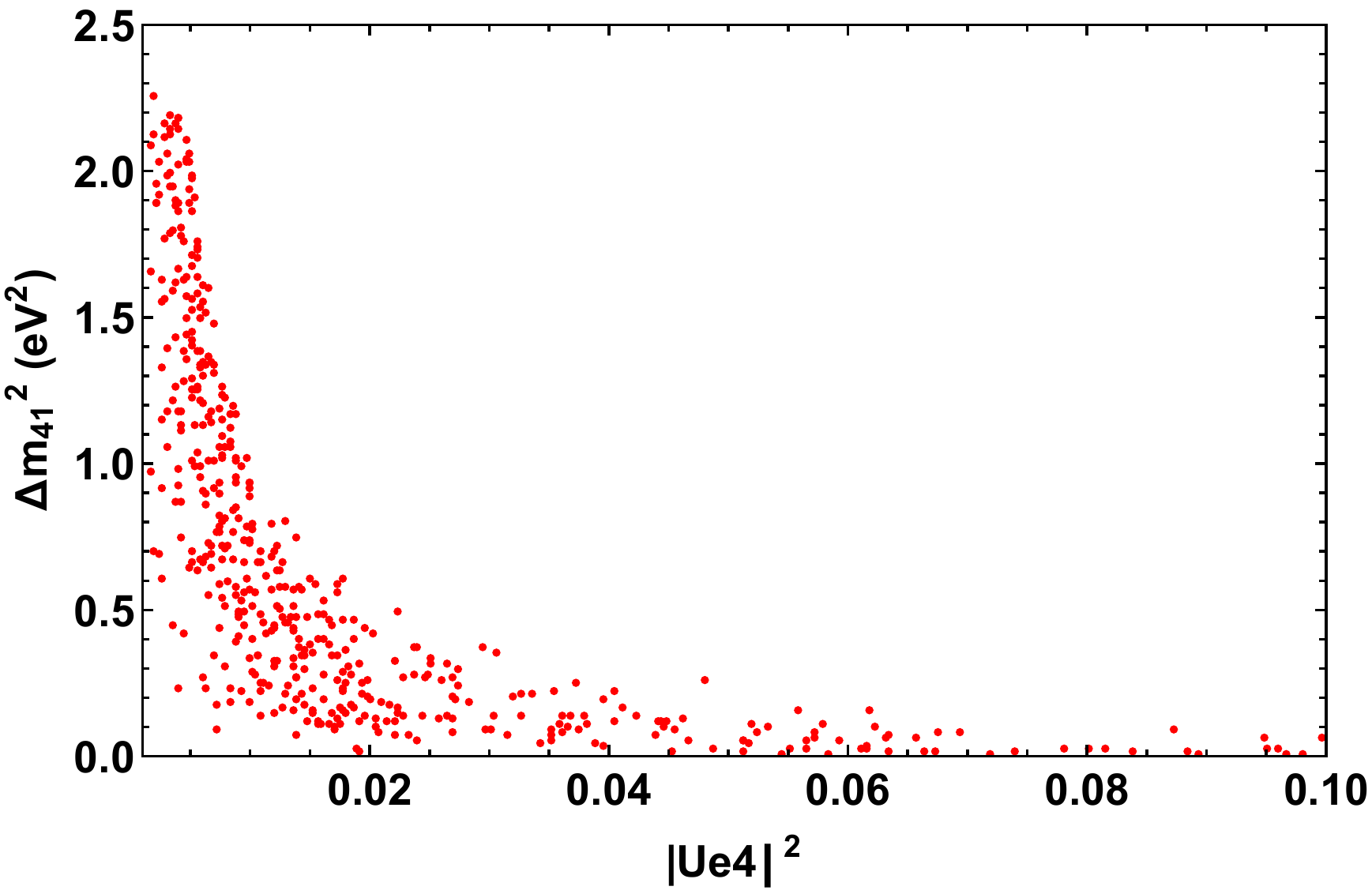}
\caption{Left panel projects the variation of model parameter \textbf{a} with  total active neutrino mass with  and  right panel represents the variation of $\phi_2$ with  total active neutrino mass.}
\label{a_mt}
\end{center}
\end{figure}

\begin{figure}[t!]
\begin{center}
\includegraphics[width=70mm,height=50mm]{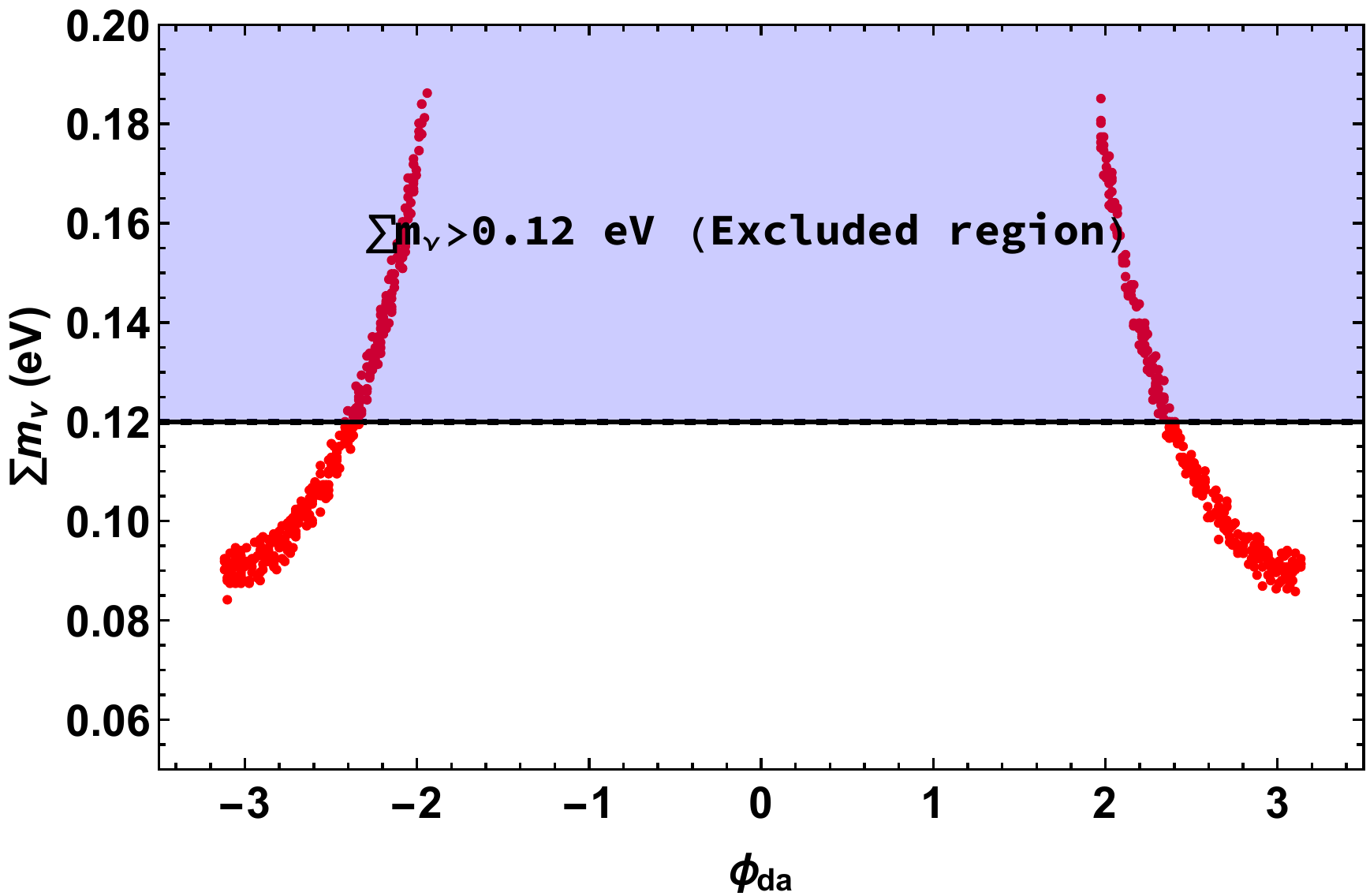}
\hspace*{0.1 true cm}
\includegraphics[width=70mm,height=50mm]{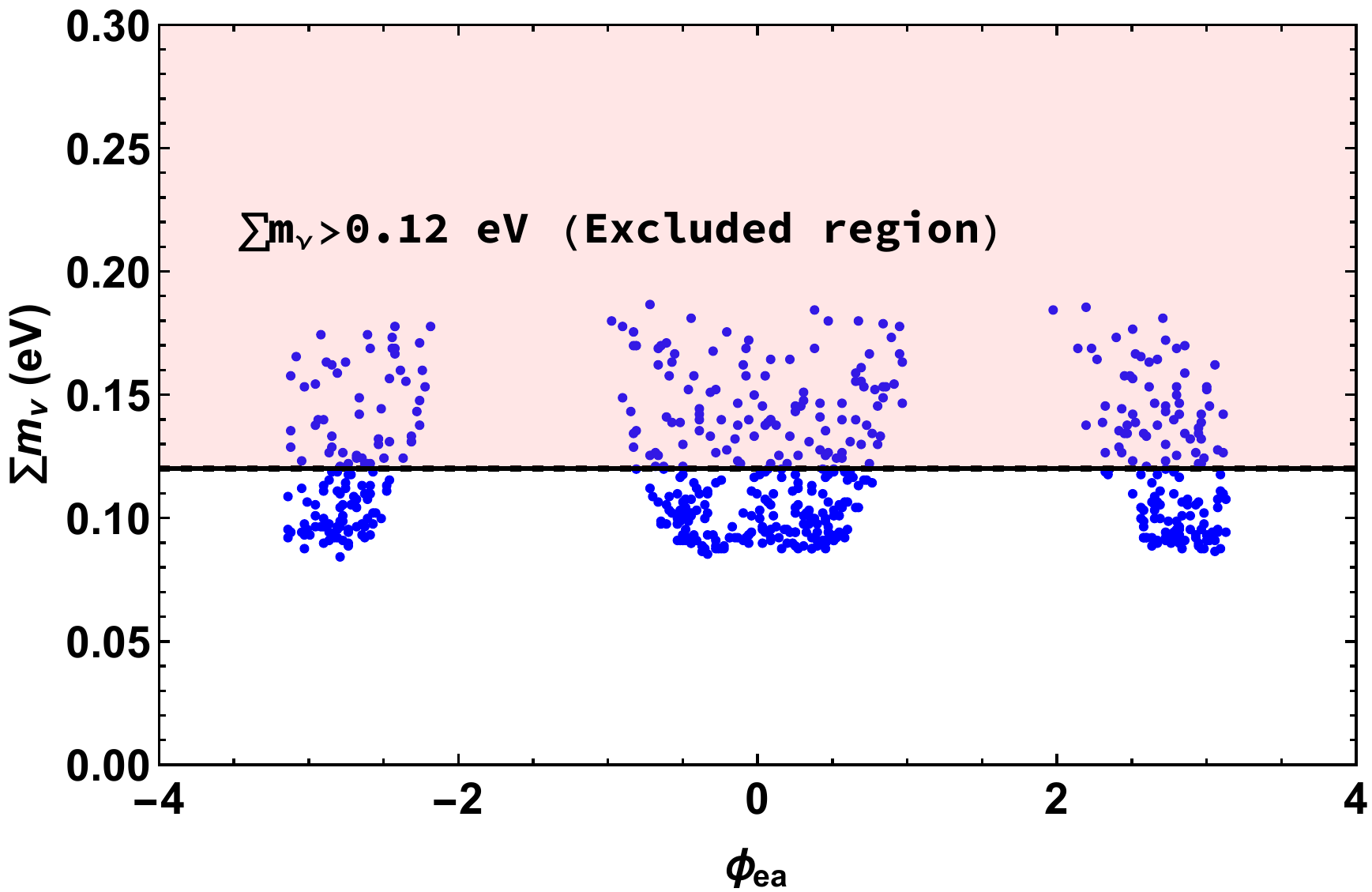}
\caption{Left (Right) panel represents the variation of $\phi_{da}$ ($\phi_{ea}$) with total active neutrino mass.}
\label{pda_pea_mt}
\end{center}
\end{figure}

We discuss  the dependence of various model parameters, which are consistent with the  $3 \sigma$ allowed ranges  of neutrino oscillation observables. The correlation and constraints on these parameters are presented in Fig. \ref{l1_s13_l3mt} to  Fig. \ref{pda_pea_mt}. Here, we fix $\lambda_2 = 0.5$ and the phase associated with $\lambda_1$, $\phi_{ba}=0$. We vary $\lambda_1$ from 0.01 to $0.3$, which in turn gives a favorable parameter space for  $\lambda_1$ to lie within $0.25$ to $0.3$, allowed by the $3 \sigma$ observation of $\theta_{13}$, which is more stringent than the constraint from total active neutrino mass as shown in the left panel of Fig. \ref{l1_s13_l3mt}. From the right panel, the allowed region for $|\lambda_3|$ turns out to be in the range $ 0.1$ to $ 0.6$. We found Majorana like phases  $\phi_1$, $\phi_2$ and  $\phi_3$ to have the allowed values  of $[-0.18 ,0.18] \pi$ (top-left panel),  $[-0.5,0.5]\pi$
(top-right panel) and $[-0.09,0.09] \pi$   (bottom panel) of Fig. \ref{p1p3_mt}.  Similarly, the left panel of Fig. \ref{a_mt} represents a strong constraint on the parameter \textbf{a} from cosmological observation of total active neutrino mass, which would lie within a range of $\pm 0.035$ to $\pm 0.05$ eV and the correlation of $|U_{e4}|^2$ with $\Delta{m^2_{41}}$ (${\rm eV}^2$) is shown in the right panel. The  phases of $\lambda_2$ and $\lambda_3$, i.e., $\phi_{da}$ and $\phi_{ea}$ are strongly constrained from neutrino mass bound. These phases are found to lie in the range, $\pm 2.4$ to $\pm 3.14$ and ($-1$ to $+1$ and $\pm 2.5$ to $\pm 3.14 $) radians respectively, as shown in the left and right panels of Fig. \ref{pda_pea_mt}. 
In the present case, by fixing $\phi_{ba} = 0$, one can have a vanishing $\delta_{13}$, even though $\theta_{13}$ remains non-zero as seen from Eq. \eqref{delta_cp}. 
\newpage
\subsubsection{Variation of model parameters by fixing $\lambda_2=0.5$ and $\phi_{ba} \neq 0$}
\begin{figure}[t!]
\begin{center}
\includegraphics[width=70mm,height=50mm]{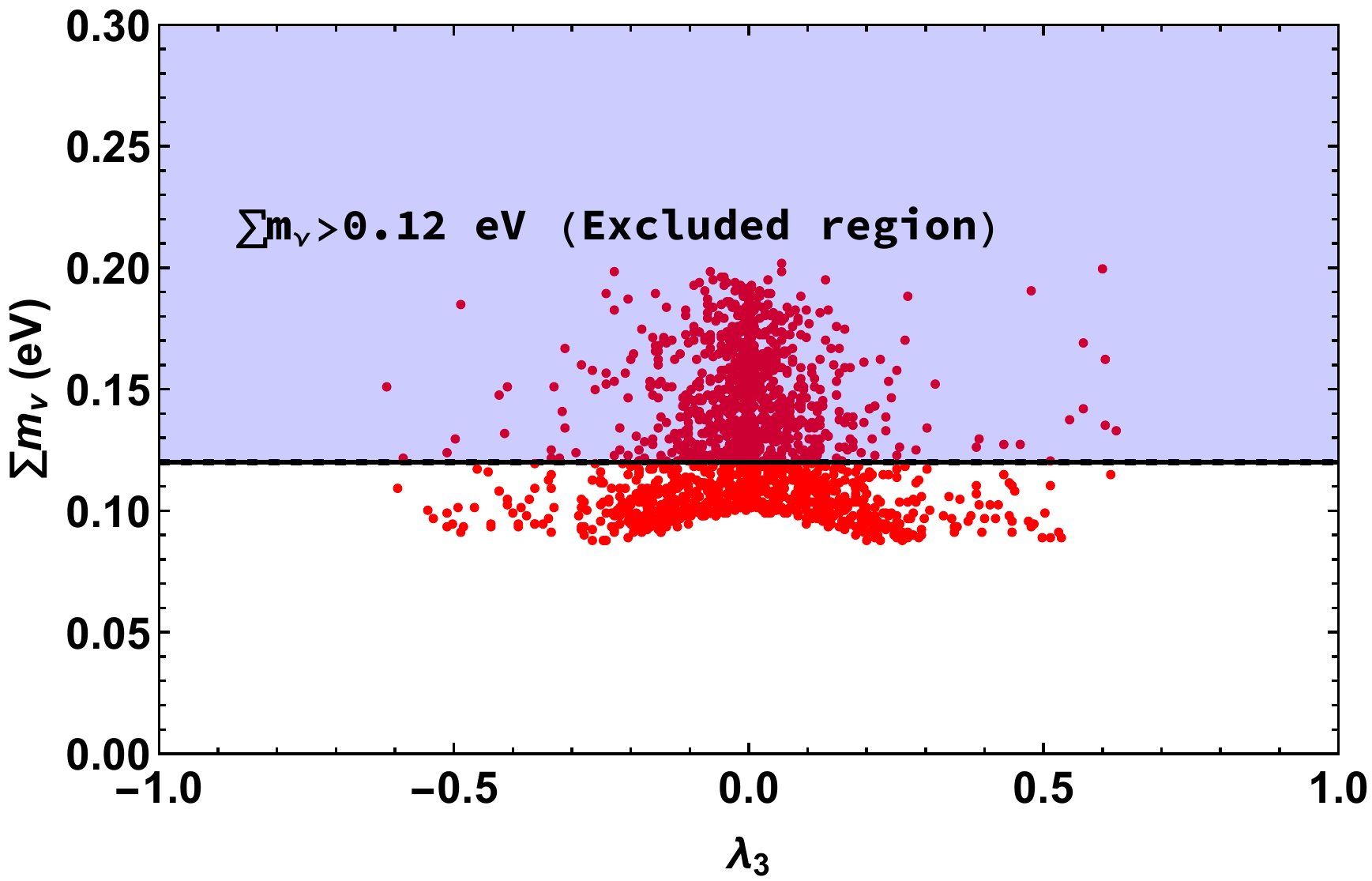}
\includegraphics[width=70mm,height=50mm]{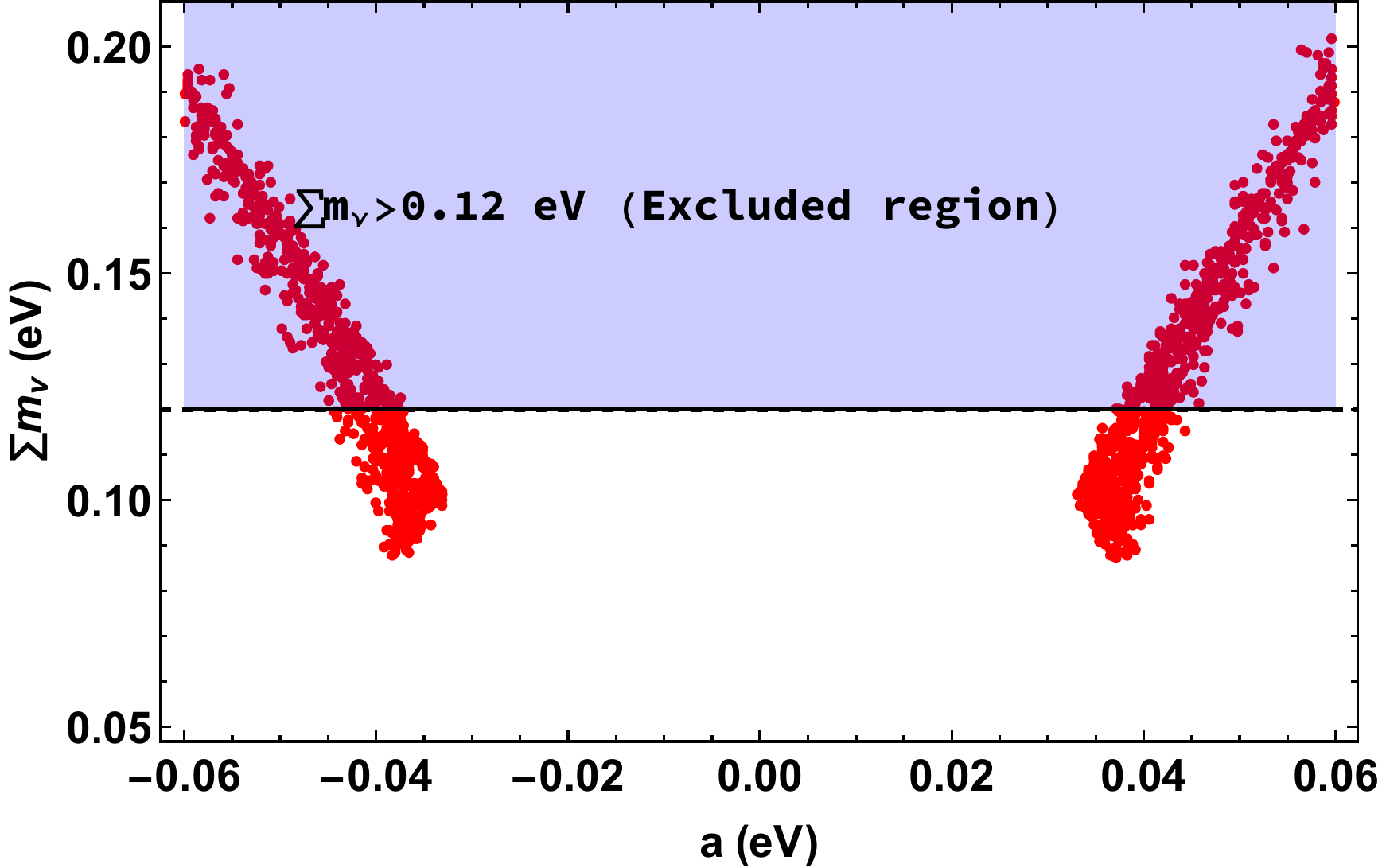}
\caption{Left (Right) panel shows the variation of $\lambda_3$ (\textbf{a}) with total active neutrino masses.}
\label{l3_mt_cp}
\end{center}
\end{figure}

\begin{figure}[t!]
\begin{center}
\includegraphics[width=70mm,height=50mm]{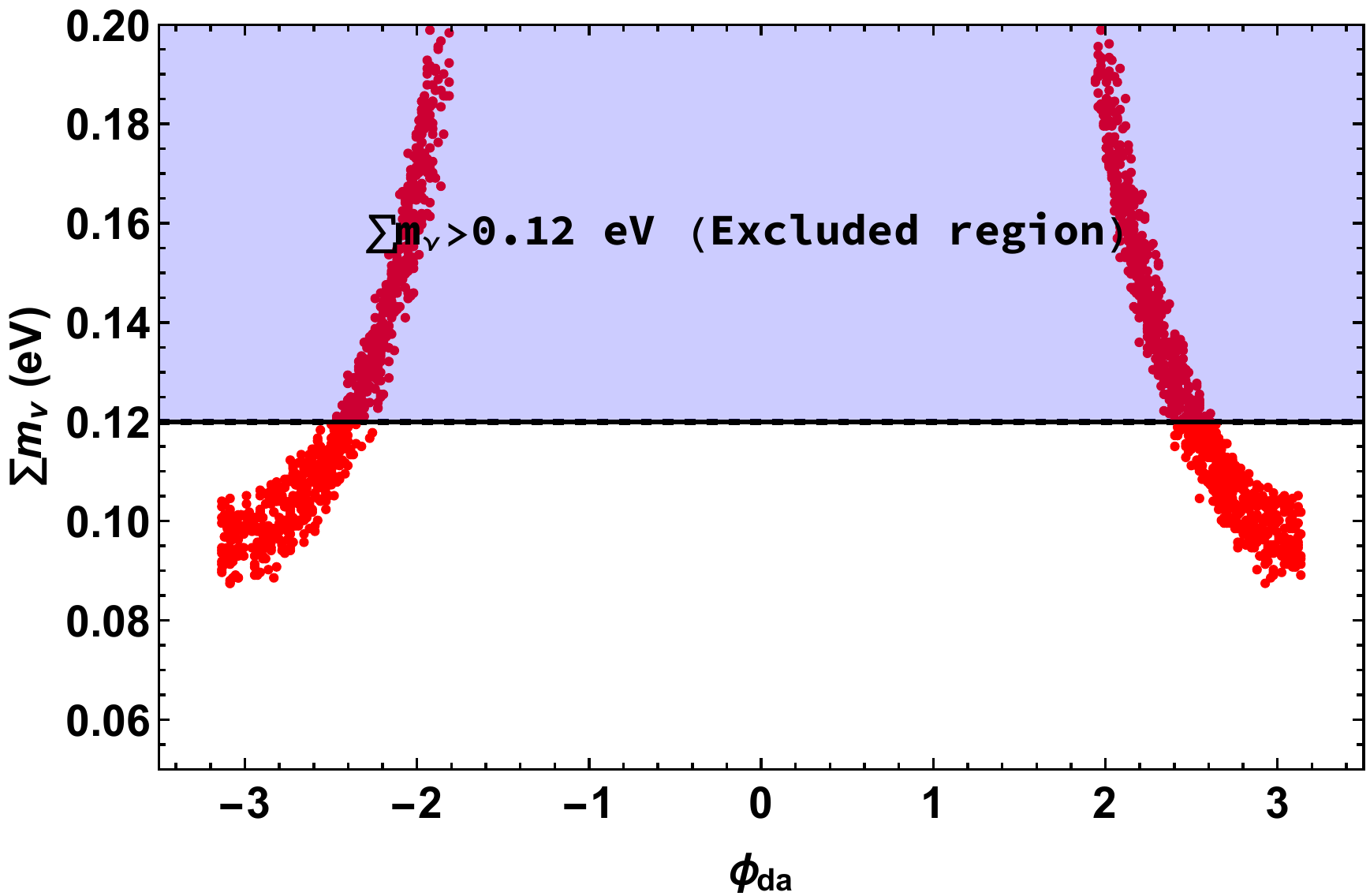}
\includegraphics[width=70mm,height=50mm]{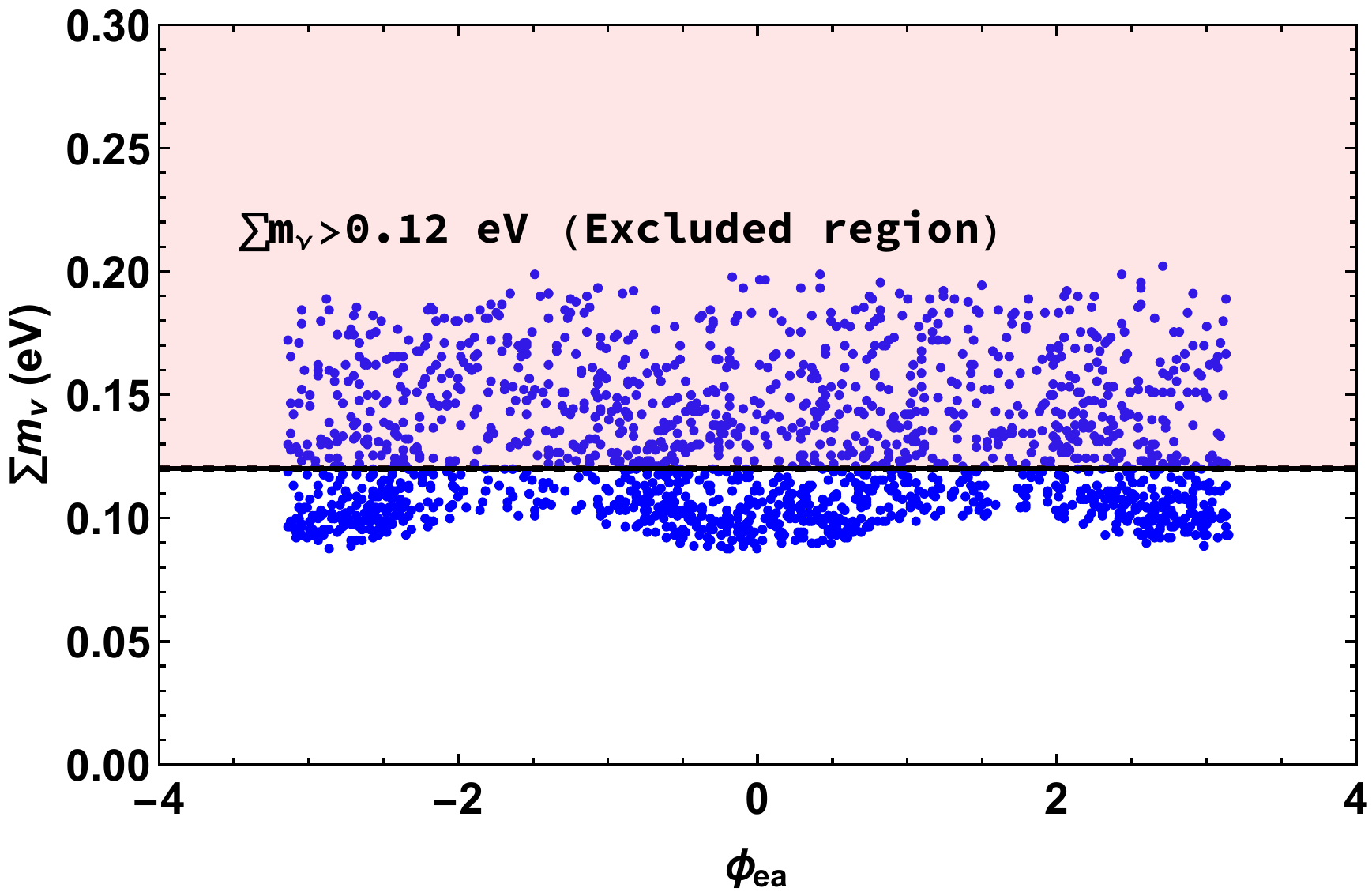}
\caption{Left (Right)  panel shows the variation of $\phi_{da}$ ($\phi_{ea}$) with total active neutrino masses.}
\label{pda_pea_cp}
\end{center}
\end{figure}

\begin{figure}[t!]
\begin{center}
\includegraphics[width=70mm,height=50mm]{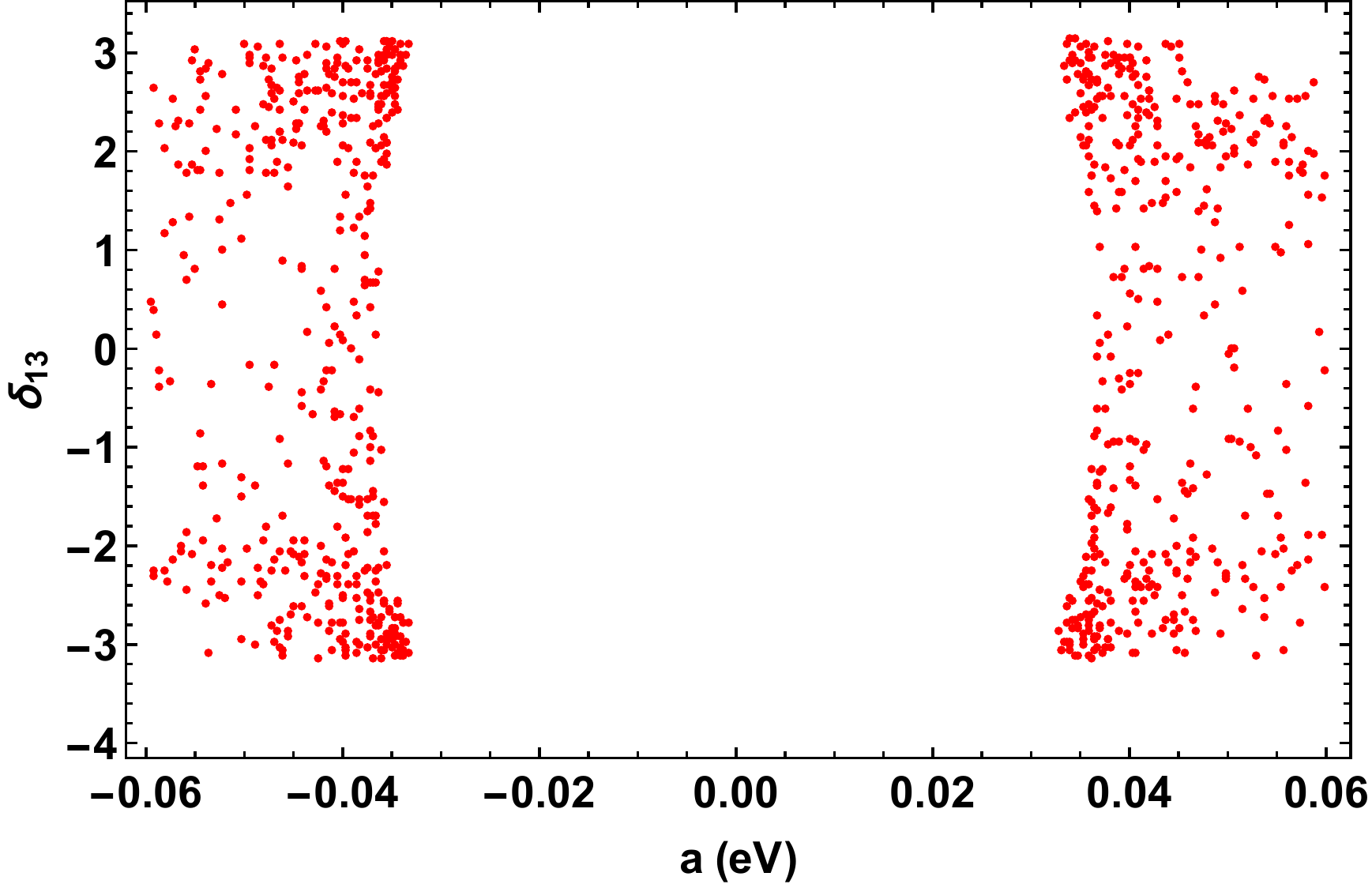}
\includegraphics[width=70mm,height=50mm]{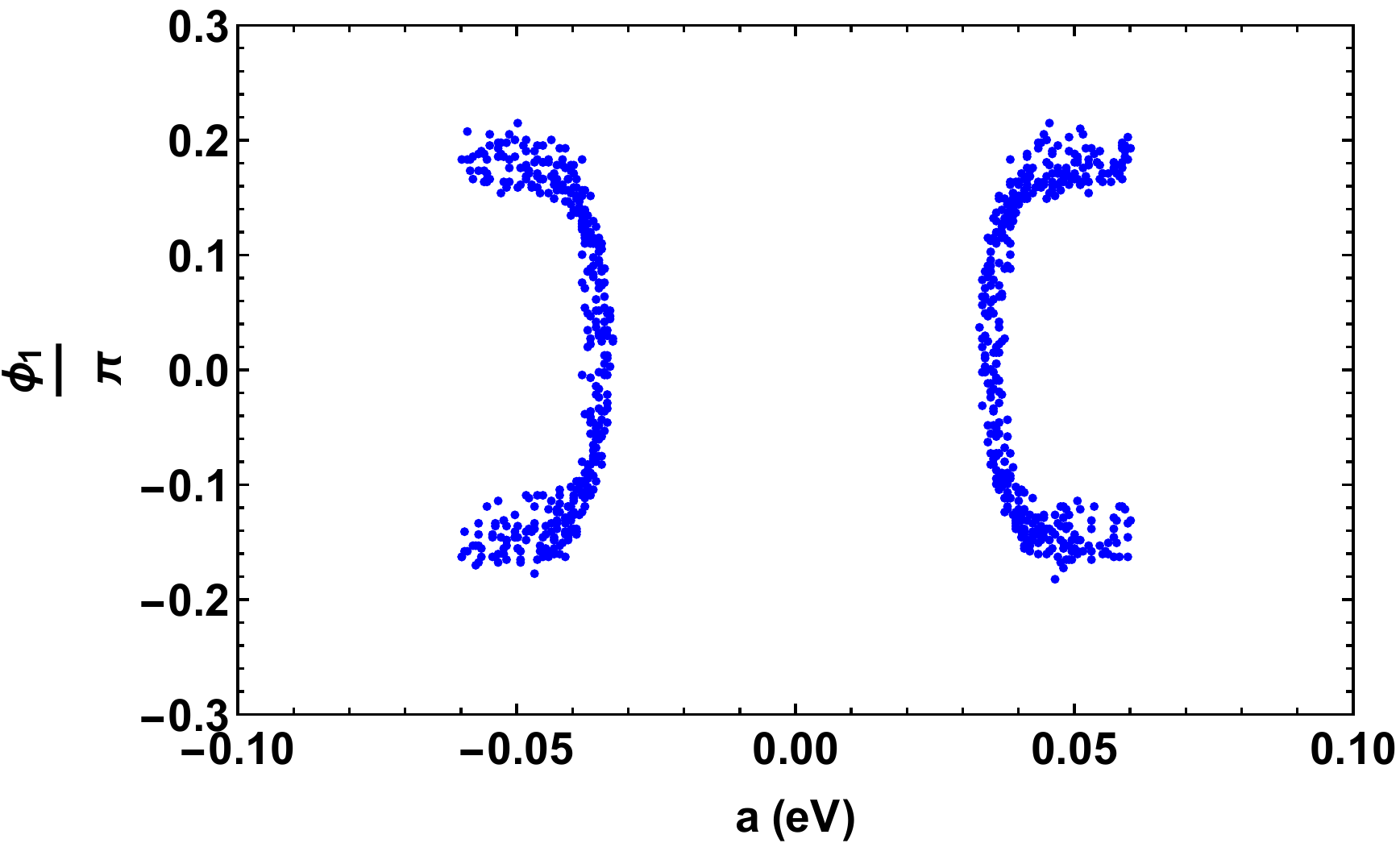}
\caption{Left  panel shows the variation of \textbf{a} with the Dirac like CP phase and the right panel depicts the correlation between \textbf{a} and $\phi_1$.}
\label{a_mt_cp}
\end{center}
\end{figure}

In the previous case, we have   a vanishing CP phase ($\delta_{13}=-\phi_{ba}$), here, we try to show the impact of non-zero $\delta_{13}$ on the model parameters. We consider the  phase of $\lambda_1$, $\phi_{ba}$ to vary from $-\pi$ to $\pi$, which changes the allowed region of model parameters, described in the previous case. From the left panel of Fig. \ref{l3_mt_cp}, we found that the region  $-0.6<\lambda_3<0.6$ is allowed by the cosmological bound on sum of active neutrino masses and the right panel shows the constraint on $\textbf{a}$ which is $\pm[0.03,0.045]$,  slightly more stringent than the previous case.. The parameter scan for $\phi_{da}$ (in radian) to lie within the domain $\pm 2.4$ to $\pm 3.14$ as shown in the left panel of Fig. \ref{pda_pea_cp}. The favored parameter space for $\phi_{ea}$ is represented in the right panel, which allows the values of $-3.14$ to $3.14$ (in radians). 
The correlation of CP phase $\delta_{13}$ and $\phi_1$ with parameter \textbf{a} are shown in the left and right panels of Fig. \ref{a_mt_cp}. In this case, we found that the other parameters do not change appreciably in comparison with the previous case.
\section{Discussion on neutrinoless double beta decay with eV-scale neutrinos.}
\label{sec:onubb}
\noindent
To accommodate an eV scale sterile-like neutrino $N_1$, with lepton number violating (LNV) Majorana mass manifest the new physics contribution beyond SM \cite{Giunti:2015kza,Abada:2018qok,Giunti:2019aiy,Huang:2019qvq}. 
The well known process of NDBD includes the simultaneous decay of two neutrons from the nucleus of an 
isotope $(A,~Z)$ into two protons and two electrons without the emission of any neutrinos in the final state \cite{Nucciotti:2007jk,Robertson:2013ziv,Cardani:2018lje,Dolinski:2019nrj,Henning:2016fad},
\begin{equation}
	\nonumber
	(A, \, Z) \rightarrow (A, \, Z + 2)^{++} + 2 e^- 
	\, . 
\end{equation}
The half-life for a given isotope can be expressed in terms of  phase-space factor $\mathcal{G}_{(A, \, Z)}^{0 \nu}$, nuclear matrix element 
$\mathcal{M}_{(A, \, Z)}^{0 \nu}$ (presented in Table.\ref{tab:nucl-matrix}) and dimensionless effective parameter $\eta_{\, \text{eff}}^{0 \nu}$, which can be inferred from the following expression, 
\begin{equation}
	(T_{1/2}^{0 \nu})^{-1}_{(A, \, Z)} 
	= 
	\mathcal{G}_{(A, \, Z)}^{0 \nu} |\mathcal{M}_{(A, \, Z)}^{0 \nu} \eta_{\, \text{eff}}^{0 \nu}|^2
	\, .
\label{halflife_formula_general}
\end{equation}
 
The experimental observation of $0\nu\beta\beta$ process will indicate the existence of an (effective) 
LNV operator. We discuss here the standard mechanism and new physics contribution to this rare process 
in the present framework with eV scale sterile-like neutrino.
 
\begin{figure}[t!]
\centering
\includegraphics[width=0.7\textwidth]{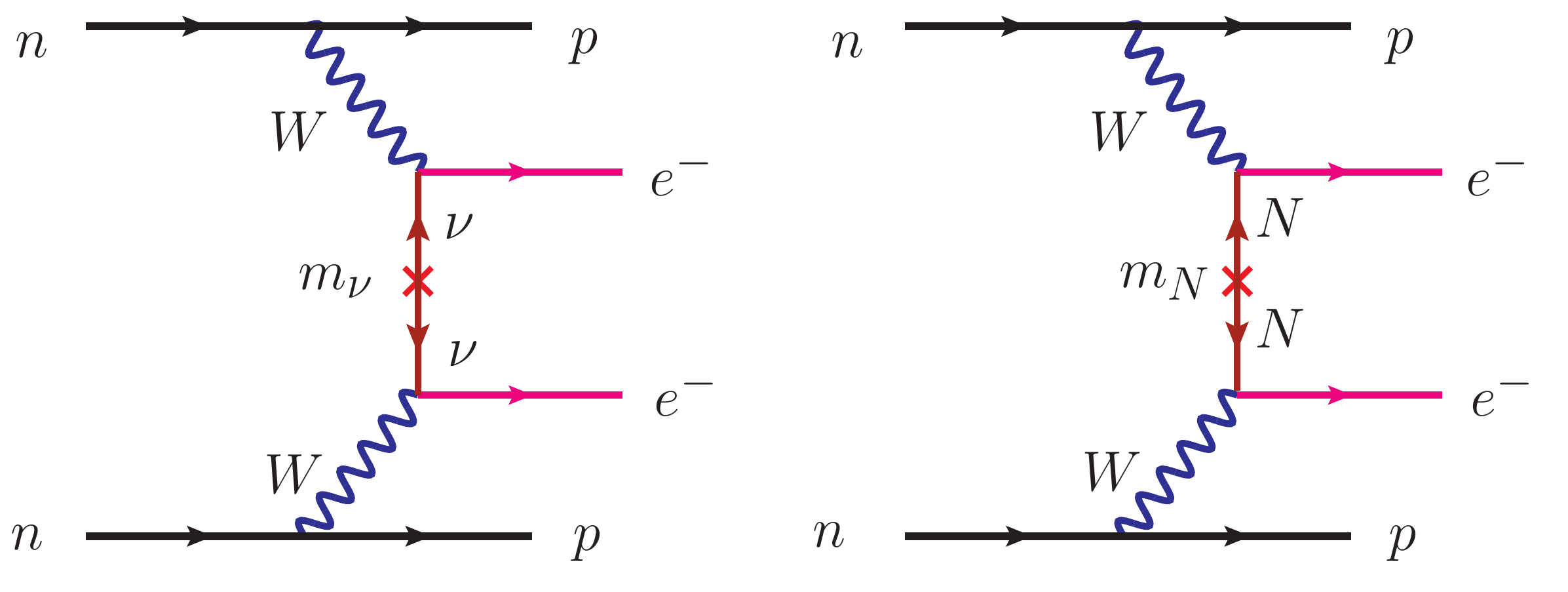}
\caption{Feynman diagrams contributing to neutrinoless double beta decay with $W^- - W^-$ mediation via the exchange of 
virtual light neutrinos $\nu$ (left panel), and the exchange of virtual eV scale sterile-like neutrinos $N$ (right panel).}
\label{fig:0nubb-feyn}
\end{figure}

\begin{table}
\centering
\vspace{10pt}
\begin{tabular}{lccc}
	\hline 
	\hline 
	Isotope	& $G_{0\nu}[10^{-15} \, {\rm yrs}^{-1}]$	& {$\mathcal{M}_\nu$}		\\
	\hline
	$^{76}$Ge	& $7.98$		& 3.85--5.82			\\ 
	$^{136}$Xe	& $59.2$		& 2.19--3.36		\\ 
\hline \hline
\end{tabular}
\caption{The numerical values of the phase-space factor and nuclear matrix elements}
\label{tab:nucl-matrix}
\end{table}

The process mediated by SM light neutrinos, which are Majorana in nature, is shown in Fig. \ref{fig:0nubb-feyn} (left panel). 
The dimensionless parameter, responsible for LNV is given by the $ee$ element of the Majorana mass matrix, normalized by the electron mass as,
\begin{equation}
	\eta_{\, \text{eff}}^{0 \nu} 
	\equiv 
	\frac{m^{\nu}_{ee}}{m_e} 
	= 
	\frac{1}{m_e} \left(	
	\sum_{i = 1}^3 (U_\text{PMNS})_{e i}^2 \, m_i
	\right)
	\, .
	\label{mee}
\end{equation}
where $U_\text{PMNS}$ is the unitary PMNS mixing matrix and $m_i$ is the mass eigenvalues of the light active neutrinos.\\
The parametrisation of the PMNS mixing matrix $U_\text{PMNS}$ is given by
\begin{eqnarray}
	U_\text{PMNS}
	=
	\begin{pmatrix}
	1 & 0 & 0 \\
	0 & c_{23} & s_{23} \\
	0 & -s_{23} & c_{23}
	\end{pmatrix}
	\begin{pmatrix}
	c_{13} & 0 & s_{13}e^{-i\delta} \\
	0 & 1 & 0 \\
	-s_{13}e^{i\delta}& 0 & c_{13}
		\end{pmatrix}
	\begin{pmatrix}
	c_{12} & s_{12} & 0 \\
	-s_{12} & c_{12} & 0 \\
	0 & 0 & 1
	\end{pmatrix}
   \begin{pmatrix}
	1 & 0 & 0 \\
	0 & e^\frac{i\alpha}{2} & 0 \\
	0 & 0 & e^\frac{i\beta}{2}
	\end{pmatrix}
	\end{eqnarray}
where $c_{ij}=\cos{\theta_{ij}} , s_{ij}=\sin{\theta_{ij}}$ are sine and cosine of the mixing angles, $\delta$ is the Dirac CP-phase and $\alpha,\beta$ are Majorana phases. 
Using the above mixing matrix $U_\text{PMNS}$ in eq.(\ref{mee}), the modified effective Majorana mass can be read as,
\begin{equation}\label{z}
 m^{\nu}_{ee}=\big | m_1 c_{12}^2 c_{13}^2 + m_2 s_{12}^2 c_{13}^2 e^{i\alpha} + m_3 s_{13}^2 e^{i\beta} \big|
\end{equation}

The effective Majorana mass depends upon the neutrino oscillation parameter $\theta_{12}$, $\theta_{13}$ and the neutrino mass eigenvalues $m_1,m_2$ and $m_3$ and phases. 
However, we do not know the absolute value of these light neutrino masses but neutrino oscillation experiments gives mass square difference between them. 
Also we have no information about these phases. In our analysis, we randomly vary these phases and took $3\sigma$ range of neutrino oscillation parameters to 
see whether we can get any crucial information about absolute scale of neutrino masses and mass ordering using experimental limit from neutrinoless double beta decay
experiments. 

We definitely know sign of $\Delta m_{sol}^2 (\equiv \Delta m_{21}^2)= m_2^2-m_1^2$ is positive which implies $m_2>m_1$. But neutrino oscillation experiments 
can not provide unambiguous sign of $m_{atm}^2(\Delta m_{31}^2)$ which allows two possible ordering of neutrino mass as,
\begin{eqnarray}
\Delta m_{atm}^2(\Delta m_{31}^2) &=& m_3^2-m_1^2 ,  ~  \textrm{for Normal Hierarchy(NH)}\nonumber\\
 &=& m_1^2-m_3^2,  ~  \textrm{for Inverted Hierarchy (IH)}\nonumber
\end{eqnarray}
\underline{\textbf{ Normal Hierarchy (NH)}} : $m_1<m_2\ll m_3$
\begin{eqnarray}
\textrm{Here},\: m_1=m_{\rm lightest} \:;\: m_2=\sqrt{m_1^2+\Delta m_{sol}^2}\, , \nonumber\\  m_3=\sqrt{m_1^2+\Delta m_{sol}^2+\Delta m_{atm}^2}
\end{eqnarray}
\underline{\textbf{ Inverted Hierarchy (IH)}} : $m_3\ll m_1<m_2$
\begin{eqnarray}
\textrm{Here}, \: m_3=m_{\rm lightest} \:;\:  m_1=\sqrt{m_3^2+\Delta m_{atm}^2} \, , \nonumber\\  m_2=\sqrt{m_3^2+\Delta m_{atm}^2+\Delta m_{sol}^2}
\end{eqnarray}

\begin{figure}
\begin{center}
\includegraphics[height=60mm,width=75mm]{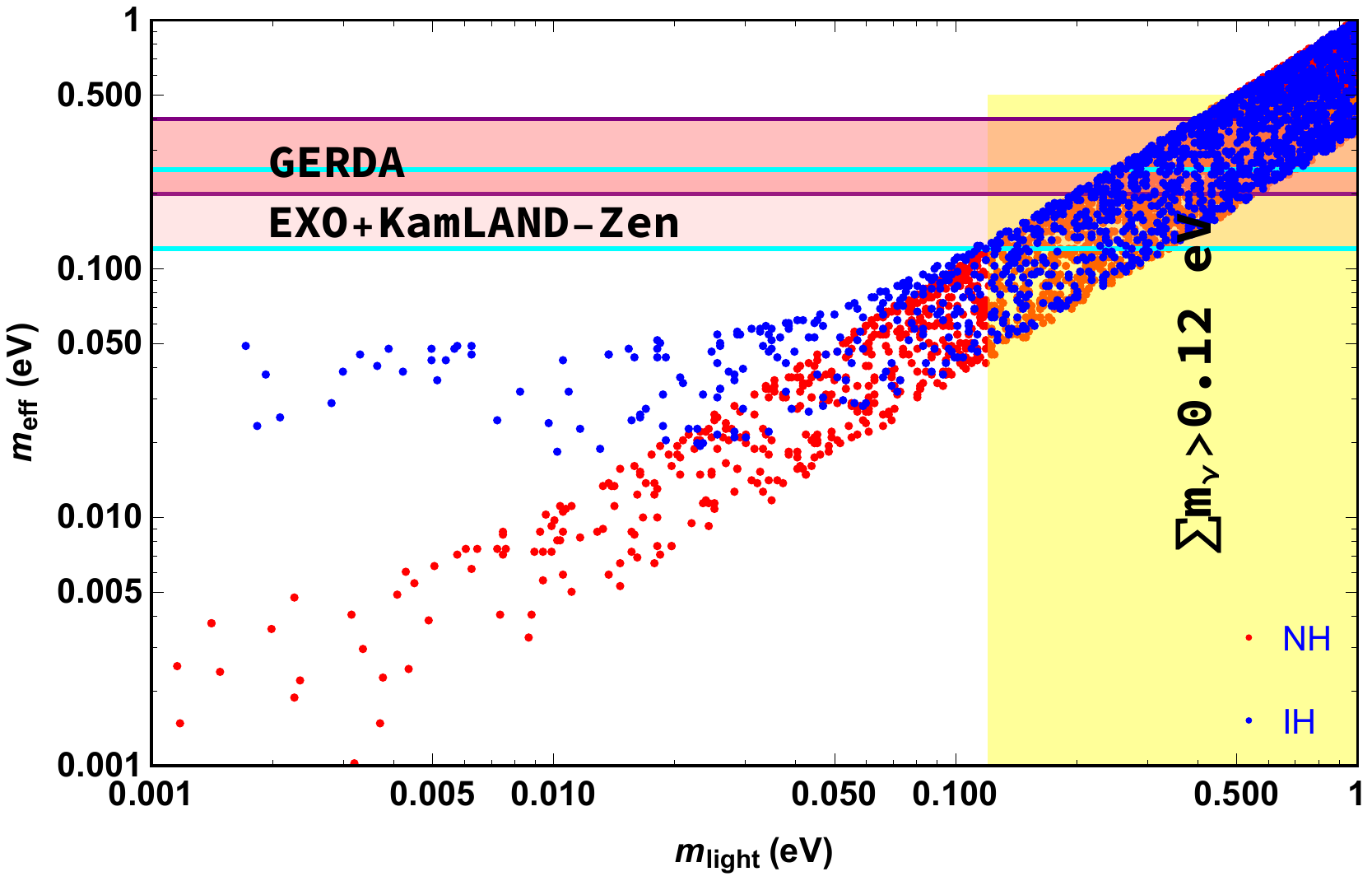}
\includegraphics[height=60mm,width=75mm]{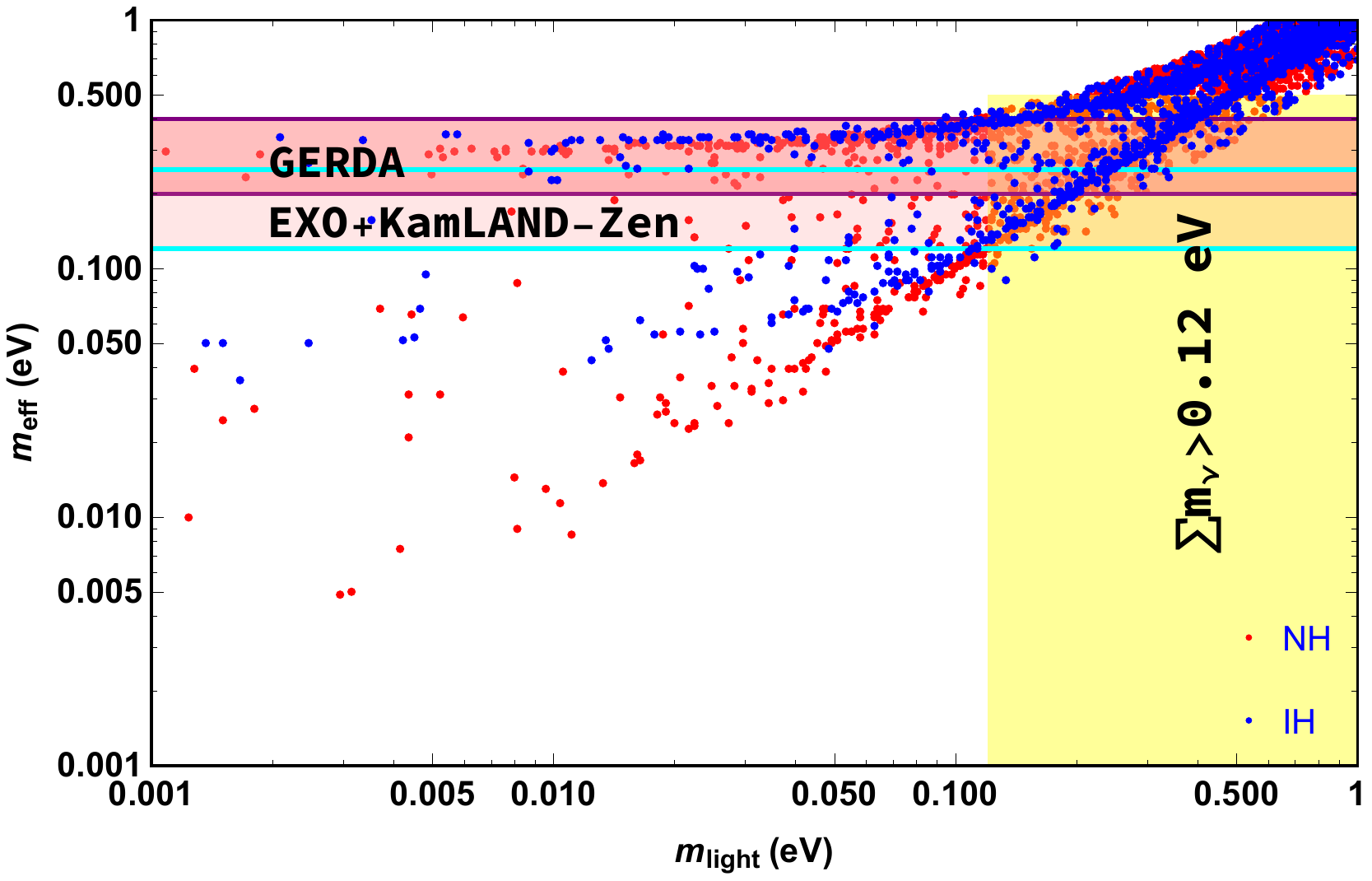}
\caption{Left Panel: Variation of effective Majorana mass as a function of the 
lightest neutrino mass, m$_1$ (m$_3$) for NH (IH) due to standard mechanism 
through light active Majoana neutrinos. Right Panel: The new physics contributions (in the presence of eV scale sterile-like neutrino, that falls in the experimental bound discussed in Table \ref{tab:0nbb_limits})
to $0\nu\beta\beta$ vs lightest neutrino mass, m$_1$ (m$_3$) for NH (IH). The NH contributions are 
displayed by red dots band while the IH contributions are given by blue dots. 
The vertical shaded area is for constraint on the sum of light neutrino masses from recent cosmological data (PLANCK1 and PLANCK2). 
The horizontal shaded areas are for the 
limits in effective Majorana mass parameter and half-life by GERDA and EXO+KamLAND-Zen experiments.  }
\label{nldbd:mee}
\end{center}
\end{figure}

Using these randomly generated input parameters and oscillation data, variations of effective mass with the lightest neutrino mass is displayed 
in Fig.\ref{nldbd:mee}. For comparison, we have taken the current experimental limits on the half-life and the corresponding mass parameter for the 
isotopes $^{76}$Ge and $^{136}$Xe as follows,
\begin{table}[h!]
\begin{center}
	\begin{tabular}{c|c|c|c}
	Isotope & $T_{1/2}^{0 \nu}~[10^{25}\text{ yrs}]$ & $m_\text{eff}^{0 \nu}~[\text{eV}]$ & Collaboration \\
	\hline
	$^{76}$Ge	& $> 2.1$	& $< (0.2 - 0.4)$ 	& GERDA~\cite{Agostini:2013mzu} 			\\
	$^{136}$Xe	& $> 1.6$	& $< (0.14 - 0.38)$ 	& EXO~\cite{Auger:2012ar} 				\\
	$^{136}$Xe	& $> 1.9$	& n/a 				& KamLAND-Zen~\cite{Gando:2012zm} 		\\
	$^{136}$Xe	& $> 3.6$	& $< (0.12 - 0.25)$ 	& EXO + KamLAND-Zen combined~\cite{Gando:2012zm} 	
	\end{tabular}
\caption{The current lower limits on the half-life $T_{1/2}^{0 \nu}$ and upper limits on the effective mass parameter 
$m_\text{eff}^{0 \nu}$ of neutrinoless double beta decay for the isotopes $^{76}$Ge and $^{136}$Xe. The range for the 
effective mass parameter comes from different calculation methods for the nuclear matrix elements.}
\label{tab:0nbb_limits}
\end{center}
\end{table}

From left-panel of Fig.\ref{nldbd:mee} and using bounds from neutrinoless double beta decay experiments and cosmology, it is quite clear that the standard mechanism is not the only way to 
realize $0\nu\beta\beta$ and more importantly, NH and IH patterns are insensitive to current experimental bound while the quasi degenerate (QD) pattern is disfavoured from cosmological data 
from PLANCK1 and PLANCK2. In principle,  we should explore all possible sources of new physics that violate lepton number (effectively) by two units and can lead to $0\nu\beta\beta$.

We explicitly found that in addition to the standard mechanism, however, there is an additional contribution coming from the new eV scale 
sterile-like neutrino. In general the light ($\nu$) and sterile-like ($N$) neutrino exchange can give the corresponding effective Majorana parameter as
\begin{equation}
	m^{\rm tot}_{ee} = m^{\nu}_{ee} + m^{N}_{ee} 
	= \left(
	  \sum_{i = 1}^3 U_{e i}^2 m_i 
	+ \sum_{i \in \text{eV}} U_{e 4}^2 m_{si}
	\right), 
\end{equation}
where, $U_{e 4}$ is the mixing between eV scale sterile neutrino with light active neutrino which has already 
expressed in terms of model parameters. 

The variation of effective Majorana mass parameter in the presence of an additional eV scale  sterile neutrino with the lightest neutrino mass 
is displayed in right-panel of Fig.\ref{nldbd:mee}. From this plot, we conclude that presence of an additional eV scale sterile-like neutrino 
enhances the predictability of the model by contributing to the NDBD (shown in Feynman diagram in Fig. \ref{fig:0nubb-feyn}). 
The dotted points which lies in the horizontal bands shows that this new physics contribution can saturate the experimental bounds from GERDA and KAMLAND on effective neutrino mass \cite{deSalas:2018bym,Penedo:2018kpc} 
and can shed light on lepton number violation in nature along-with its implication to cosmology like matter-antimatter asymmetry of the universe.


\section{Dark Matter Phenomenology}
\begin{figure}[t!]
\begin{center}
\includegraphics[width=0.3\linewidth]{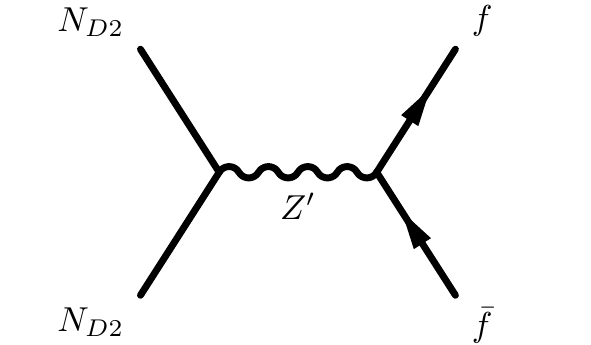}
\includegraphics[width=0.3\linewidth]{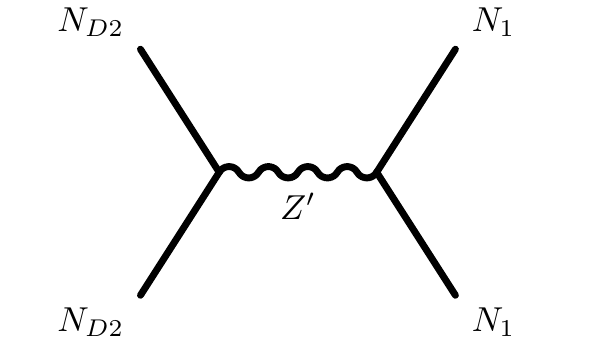}
\includegraphics[width=0.3\linewidth]{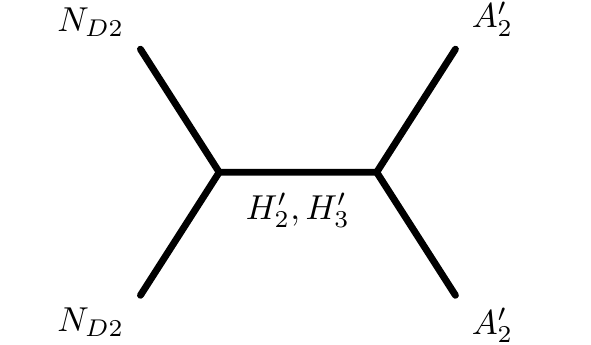}
\caption{Annihilation channels for $N_{D2}$ contributing to relic density and similar diagrams will be followed for $N_{D3}$.}
\label{relic_feyn}
\end{center}
\end{figure}

\begin{figure}[t!]
\begin{center}
\includegraphics[width=0.48\linewidth]{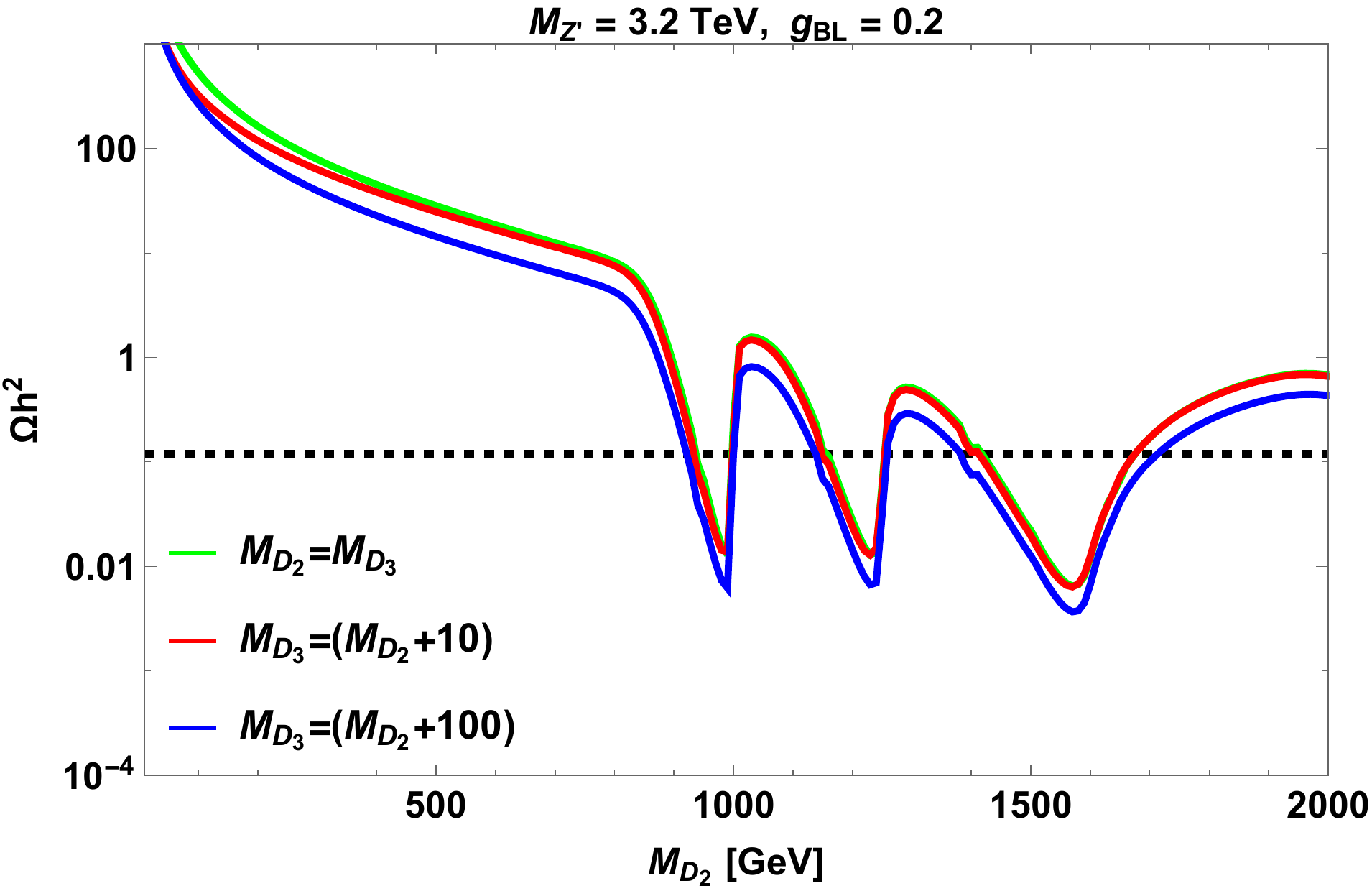}
\vspace{0.2 cm}
\includegraphics[width=0.48\linewidth]{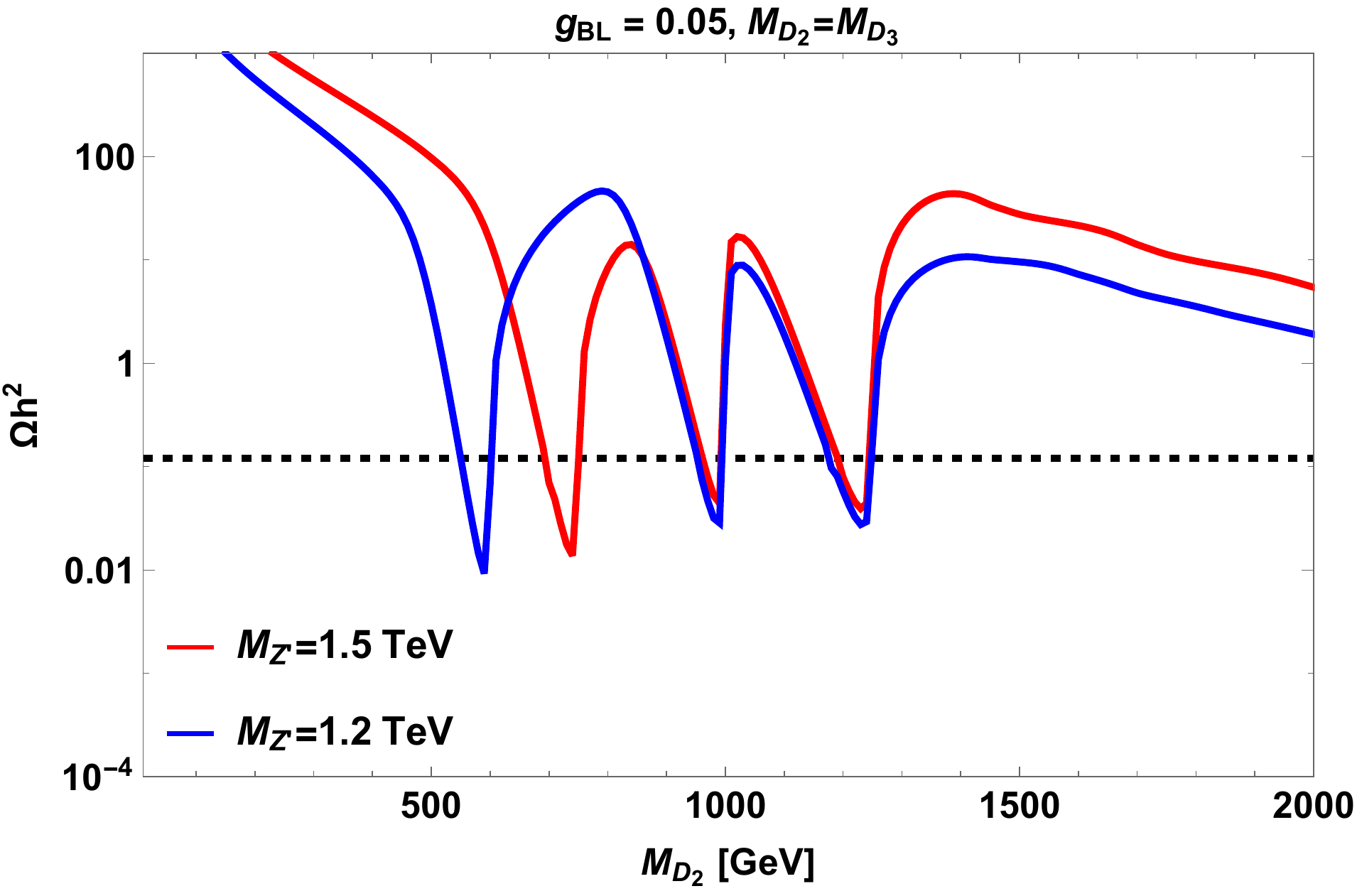}
\caption{Variation of relic density as a function of DM mass ($M_{D_2}$) with fixed $Z^\prime$ mass and $g_{\rm BL}$  for different set of mass splittings between $M_{D_2}$ and $M_{D_3}$. Right panel corresponds to the behavior of relic density by varying $Z^\prime$ mass for $M_{D_2} = M_{D_3}$. The benchmark for the masses 
of the scalars are $(M_{H^\prime_{1}},M_{H^\prime_{2}},M_{H^\prime_{3}},M_{A^\prime_{1}},M_{A^\prime_{2}}) = (2.2,2,2.5,2.1,0.9)$ (in TeV). Horizontal dashed 
lines represent $3\sigma$ range of Planck limit on relic density.}
\label{relic_onesterile}
\end{center}
\end{figure}

The model includes two heavy Majorana neutrinos $N_2$ and $N_3$ with exotic $\rm B-L$ charge $-4$, which ensures their stability by forbidding the interaction with SM particles, and the  interaction Lagrangian is given as 
\begin{eqnarray}
\mathcal{L}_{\rm DM}=\sum_{\alpha=2,3}i\overline{N_{\alpha}}\gamma^\mu D_\mu N_\alpha-\sum_{\alpha,\beta=2,3}\left[ y_{\alpha \beta} N_{\alpha} N_{\beta} \phi_8+~{\rm H.c}\right],\label{DMlag}
\end{eqnarray} 
where, $D_\mu=\partial_\mu -4ig_{\rm BL}Z^\prime_\mu.$
These fermions acquire masses when the local $B-L$ symmetry is broken and thereby leading to an inherent symmetry of $N_{2,3} \rightarrow -N_{2,3}$, which mimics the $Z_2$ symmetry, ensuring the stability of the DM \cite{Bonilla:2018ynb,Bonilla:2019hfb}. All the two body decays of these DM candidates are kinematically forbidden at re-normalizable level, since all the generic Yukawa interactions with SM leptons are not allowed and by assuming the mass of the scalar field is greater than mass of the dark matter candidates. The mixing matrix of these two neutral fermions from \eqref{DMlag} is given by
\begin{equation}
M_R=\begin{pmatrix}
y_{22} \frac{v_8}{\sqrt{2}}  && y_{23} \frac{v_8}{\sqrt{2}}\\
y_{23} \frac{v_8}{\sqrt{2}} &&  y_{33} \frac{v_8}{\sqrt{2}}
\end{pmatrix}.
\end{equation}
The above mass matrix can be diagonalized by orthogonal transformation:$ U_R M_R {U_R}^T = M_D $, where
\begin{equation}
U_R=\begin{pmatrix}
\cos{\theta} && \sin{\theta}\\
-\sin{\theta} && \cos{\theta}
\end{pmatrix}, \hspace{5mm} M_D=\begin{pmatrix}
M_{D_2} && 0\\
0 && M_{D_3}
\end{pmatrix}.
\end{equation}
Here, the mixing angle is given by $\theta=\frac{1}{2}\tan^{-1}\left[\frac{2 y_{23} v_8 }{y_{33} v_8-y_{22} v_8}\right]$. The flavor and mass eigenstates are related by
\begin{eqnarray}
&&N_2 = \cos\theta~ N_{D2}  + \sin\theta~N_{D3},\nn\\
&&N_3 = -\sin\theta ~N_{D2} + \cos\theta~N_{D3}.
\label{eigen}
\end{eqnarray}
In the present context, both of mass eigenstates $N_{D2}$, $N_{D3}$ are stable dark matter candidates as both of them do not decay and hence, the total relic is summed up ($\Omega h^2 = \Omega_2 h^2 +\Omega_3 h^2$) \cite{Bernal:2018aon}. Due to the same quantum numbers and similar mass mechanism, both of the heavy fermions will have comparable masses, but a small mass splitting can be generated by adjusting the Yukawa couplings. Using Eq. \ref{eigen} in Eq. \ref{DMlag}, one can show that the interference terms (involving $N_{D2}$ and $N_{D3}$) with the gauge boson $Z^\prime$ and the scalar $\phi_8$ vanishes. Hence, there will be no coannihilation effect in the computation of relic density. We use the well-known 
packages LanHEP \cite{Semenov:1996es} and micrOMEGAs \cite{Pukhov:1999gg, Belanger:2006is, Belanger:2008sj} for the DM analysis. Channels giving 
significant contribution to relic density are shown in Fig. \ref{relic_feyn}. Annihilation to sterile-like neutrinos ($N_1N_1$ in final state) in gauge portal and 
CP-odd scalars ($A_2^\prime A_2^\prime$ in final state) in scalar portal, stand out to give major contribution.  We have fixed the masses of the scalar spectrum 
and gave emphasis to the impact of gauge parameters $M_Z^\prime$ and $g_{\rm BL}$ on relic density.  Left panel of Fig. \ref{relic_onesterile}  depicts the 
behavior of relic density with DM mass ($M_{D_2}$) for various mass splittings between $M_{D_2}$ and $M_{D_3}$, the right panel represents the behavior for different $Z^\prime$ masses. Relic density with s-channel contribution is supposed to give resonance in 
the propagator ($Z^\prime$, $H_2^\prime$, $H_4^\prime$).

\begin{figure}[t!]
\begin{center}
\includegraphics[width=0.48\linewidth]{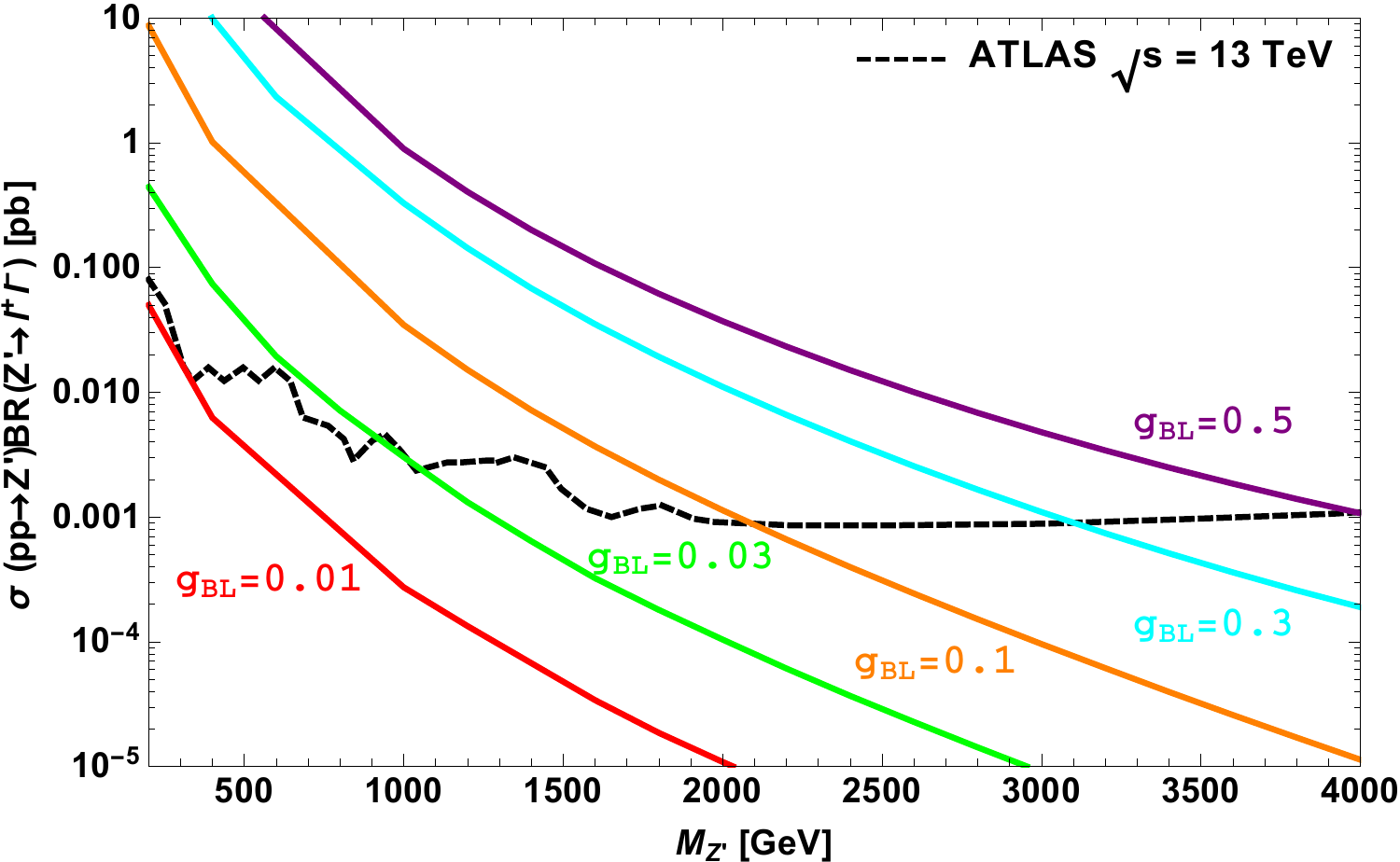}
\includegraphics[width=0.48\linewidth]{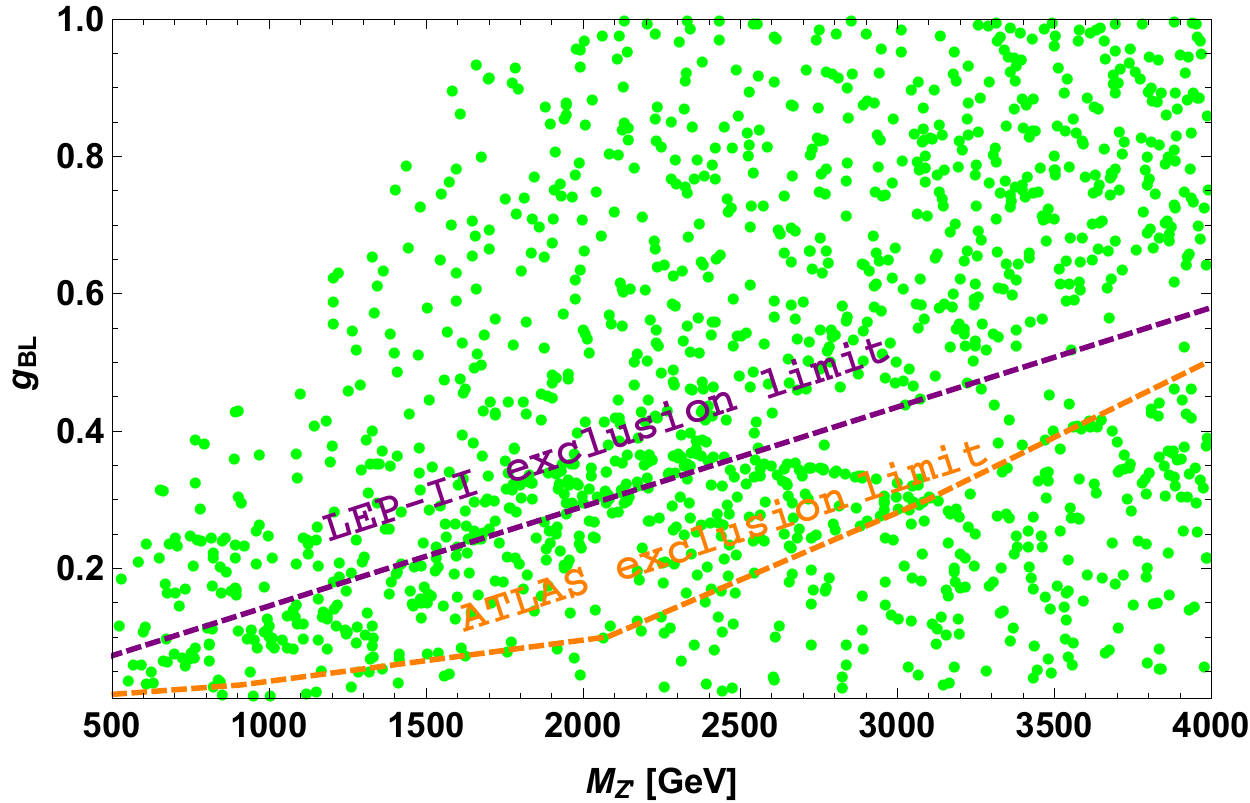}
\caption{Colored lines in left panel represent the dilepton signal cross section as a funciton of $M_{Z^\prime}$ for different values of $g_{
\rm BL}$ with the black dashed line points to ATLAS bound \cite{ATLAS-CONF-2015-070}. Right panel projects the constraint by ATLAS and LEP-II \cite{Schael:2013ita} 
on the gauge parameter space for $M_{D_2}=M_{D_3}$.}
\label{ATLAS_onesterile}
\end{center}
\end{figure}

As $Z^\prime$ couples axial vectorially with DM fermion and vectorially with SM fermion, the WIMP-nucleon cross-section is not sensitive to direct detection 
experiments. Moving to the parameter scan, the gauge parameters $M_{Z^\prime}$ and $g_{\rm BL}$ are restricted from the searches of dilepton signals in 
$Z^\prime$-portal by ATLAS \cite{ATLAS-CONF-2015-070}, and also LEP-II \cite{Schael:2013ita}. We have used CalcHEP \cite{Belyaev:2012qa,Kong:2012vg} 
to obtain the cross section $pp \to Z^\prime \to ee(\mu\mu)$ as a function $Z^\prime$ mass, depicted in the left panel of Fig. \ref{ATLAS_onesterile}. It can be 
seen that for $g_{\rm BL} = 0.01$, the region $M_{Z^\prime} < 0.3$ TeV is excluded and for $g_{\rm BL} = 0.03$, the allowed region is $M_{Z^\prime} > 0.9$ TeV. 
For $g_{\rm BL} = 0.1$, the $M_{Z^\prime}$ should be above $2$ TeV. $M_{Z^\prime}>3$ TeV is allowed for $g_{\rm BL} = 0.3$ and heavy mass regime for 
$Z^\prime$ (above $4$ TeV) is favorable for $g_{\rm BL} = 0.5$. Right panel of Fig. \ref{ATLAS_onesterile} projects the parameter space consistent with 
Planck relic density limit upto $3\sigma$ range, with the exclusion limits of ATLAS and LEP-II $\left(\frac{M_{Z^\prime}}{g_{\rm BL}} > 6.9~{\rm TeV}\right)$. 
The favorable region refers to the data points below both the experimental bounds.


\newpage
\section{Summary and Conclusion}
In this article, we have presented a detail study of neutrino and dark matter phenomenology in a minimal extension of Standard Model with $\rm U(1)_{B-L}$ and $A_{4}$ flavor symmetry. The model includes additional three neutral fermions with exotic $\rm B-L$ charges of $-4$,$-4$ and $5$ for cancellation of triangle gauge anomalies. The scalar sector is enriched with six SM singlet fields, of them, three are assigned with $\rm U(1)_{ B-L}$ charges and the rest three are charged under $A_4$ flavor symmetry. The former scalar fields helps in spontaneous breaking of $\rm B-L$ symmetry and one massless mode of the CP odd eigenstates gets absorbed by the new gauge boson $Z'$. The later scalar fields, known as $A_4$ flavons, break the $A_4$ flavor symmetry spontaneously at high scale before the breaking of $U(1)_{\rm B-L}$.

       As different short-baseline experiments such as LSND, MiniBooNE etc. are pointing towards the existence of eV scale sterile neutrinos to explain certain experimental discrepancies, we tried to address the neutrino phenomenology with a fourth generation sterile-like neutrino. Out of the three exotic fermions in the model, one is in  eV scale (sterile-like) and rest of them are in TeV scale, which help in explaining the neutrino mass and DM simultaneously. The presence of discrete symmetry provides a specific flavor structure to the neutrino mass matrix, leads to a better phenomenological consequences of neutrino mixing with a fourth generation sterile-like neutrino. We explored the active-sterile mixing in compatible with the current experimental observation. We found a  large $\theta_{13}$ and associated non-zero CP phase, within the observed $3\sigma$ range of LSND data by introducing a perturbation term to the Lagrangian. Presence of eV scale  sterile-like neutrino also provides an allowed parameter space for the effective neutrino mass in NDBD, lies within the experimental bound of KamLAND-Zen and GERDA. We strongly constrained the model parameters from the cosmological bound of active neutrino masses and showed the correlation between different mixing angles.
      
       Apart from neutrino mixing, we studied the dark matter phenomenology with rest two exotic fermions, by generating the tree level mass in TeV scale unlike the eV scale sterile-like neutrino. By introducing suitable singlet scalar, the mass mechanism for these heavy fermions is assured by the $\rm B-L$ breaking in TeV scale. Within the model framework, we found that both the Majorana fermions satisfy the correct  DM relic density and the total contribution follows the $3 \sigma$ observation of Planck, with s-channel annihilation processes in scalar and new gauge portal.  As expected, the s-channel resonances for scalars and heavy gauge boson are obtained in the relic density, we have also analyzed the behavior for different values of model parameters. We also strongly constrained the parameters associated with gauge mediated processes from the ATLAS studies of di-lepton signals and LEP II. We have shown the allowed parameter space satisfying DM and collider constraints. Direct detection of dark matter in $Z'$ portal is insensitive to direct detection experiments because of the Majorana nature. Finally, the proposed idea of extending SM with both gauge and flavor symmetries provides a suitable platform to investigate both neutrino and dark sectors. 

\acknowledgments 
Subhasmita Mishra and Mitesh Kumar Behera would like to acknowledge DST for its financial support. RM  acknowledges the support from  SERB, Government of India, through grant No. EMR/2017/001448. SM would like to acknowledge Prof. Anjan Giri for his support and useful discussion. We acknowledge the use of CMSD HPC  facility of Univ. of Hyderabad to carry out the computational work. 

\section*{Appendix}
$A_4$ symmetry includes three and one dimensional irreducible representations.\\
If $\begin{pmatrix}
a_1,a_2,a_3
\end{pmatrix}$  and $\begin{pmatrix}
b_1,b_2,b_3
\end{pmatrix}$ are the triplets of $A_4$, tensor products of these triplets are given as following
\begin{eqnarray*}
& 3\otimes 3=3_s \oplus 3_A \oplus 1  \oplus 1'\oplus 1'',\\
& 1 \otimes 1=1  , \hspace{3mm} 1' \otimes 1''=1\;,\\
& 1' \otimes 1'=1''   ,\hspace{3mm} 1'' \otimes 1''=1',\\
\\
\text{where} \hspace{4mm} & 3_s=\begin{pmatrix}
2a_1 b_1-a_2 b_3-a_3 b_2\\
2 a_3 b_3-a_1 b_2- b_1 a_2\\
2 a_2 b_2 -a_3 b_1-a_1 b_3
\end{pmatrix}       , \hspace {6mm} 3_A=\begin{pmatrix}
a_2 b_3-a_3 b_2\\
a_1 b_2-a_2 b_1\\
a_3 b_1-a_1 b_3\\
\end{pmatrix},\\
\\
& 1=a_1 b_1+a_2 b_3+a_3 b_2,\\
& 1' =a_3 b_3+a_1 b_2+a_2 b_1,\\
& 1''=a_2 b_2+a_3 b_1+a_1 b_3\;.\\
\end{eqnarray*}

\bibliography{BL}

\end{document}